\documentclass[twocolumn]{aastex631}
\usepackage{amsmath,hyperref}
\usepackage{lipsum}
\usepackage{graphicx}
\usepackage{rotating}
\accepted{March 12, 2023}

\shorttitle{Giant Impacts and Debris Disk Shapes}
\shortauthors{Jones et al.}

\begin{document}

\title{GIANT IMPACTS AND DEBRIS DISK MORPHOLOGY}

\author{Joshua W. Jones}
\affiliation{Department of Electrical Engineering and Computer Science, University of California Berkeley, 
Berkeley, CA 94720-1770, USA}

\author[0000-0002-6246-2310]{Eugene Chiang}
\affiliation{Department of Astronomy, University of California Berkeley, 
Berkeley, CA 94720-3411, USA}
\affiliation{Department of Earth and Planetary Science, 
University of California Berkeley, 
Berkeley, CA 94720-4767, USA}

\author[0000-0002-5092-6464]{Gaspard Duch\^ene}
\affiliation{Department of Astronomy, University of California Berkeley, 
Berkeley, CA 94720-3411, USA}

\author[0000-0002-6221-5360]{Paul Kalas}
\affiliation{Department of Astronomy, University of California Berkeley, 
Berkeley, CA 94720-3411, USA}
\affiliation{SETI Institute, Carl Sagan Center, 189 Bernardo Ave.,  Mountain View, CA 94043, USA}
\affiliation{Institute of Astrophysics, FORTH, GR-71110 Heraklion, Greece}

\author[0000-0002-0792-3719]{Thomas M.~Esposito}
\affiliation{Department of Astronomy, University of California Berkeley, 
Berkeley, CA 94720-3411, USA}
\affiliation{SETI Institute, Carl Sagan Center, 189 Bernardo Ave.,  Mountain View CA 94043, USA}


\begin{abstract}
Certain debris disks have non-axisymmetric shapes in scattered light which are unexplained. The appearance of a disk depends on how its constituent Keplerian ellipses are arranged. The more the ellipses align apsidally, the more non-axisymmetric the disk. Apsidal alignment is automatic for fragments released from a catastrophic collision between solid bodies. We synthesize scattered light images, and thermal emission images, of such giant impact debris. Depending on the viewing geometry, and if and how the initial apsidal alignment is perturbed, the remains of a giant impact can appear in scattered light as a one-sided or two-sided ``fork'', a lopsided ``needle'', or a set of ``double wings''. Double wings are difficult to reproduce in other scenarios involving gravitational forcing or gas drag, which do not align orbits as well. We compare our images with observations and offer a scorecard assessing whether the scattered light asymmetries in HD 15115, HD 32297, HD 61005, HD 111520, HD 106906, $\beta$ Pic, and AU Mic are best explained by giant impacts, gravitational perturbations, or sculpting by the interstellar medium.
\end{abstract}



\section{Introduction} \label{sec:intro}

Orbiting stars $\gtrsim 10$ Myr old, debris disks trace the 
last stages of planet formation \citep{hughes18}.  Their optical
thinness allows unobscured views of sub-$\mu$m to mm-sized grains 
in optical to radio-wavelength images.
It has been hoped that such images may be used to
uncover much larger objects, planets in principle, which though not
directly detectable would still reveal themselves by shaping
disks in telltale ways.

Gravity is one means by which otherwise invisible masses can change a disk's appearance. The perturbers can be planets (e.g.~\citealt{wyatt99}; \citealt{quillen06}; \citealt{stark09}) or stars (e.g.~\citealt{larwood01}; \citealt{augereau04}; \citealt{fehr22}; \citealt{farhat22}), altering the trajectories of disk grains which serve as test particles. Or the disk can perturb itself by self-gravity (e.g.~\citealt{jalali12}; \citealt{sefilian21}).

Whatever the source of the perturbation, the dominant mass remains that of the central star, and in a nearly point-mass potential, orbits of dust grains approximate Keplerian ellipses. The appearance of a disk depends not only on how elliptical the orbits are, but also on how the orbits are arranged. The more the ellipses apsidally align, the more non-axisymmetric the disk. To an extent that we expand on below, apsidal alignment of disk particles can be secularly enforced by an eccentric planet. In this scenario, depending on the viewing geometry, a variety of disk shapes can be generated in scattered light \citep{lee16}. Among these are ``needles'' which are short on one ansa and long on the other, and ``wings'' where the ansae are upturned.

Apsidal alignment can also result from a giant impact.  
The sudden destruction of a progenitor body on an initially circular orbit creates debris with a range of orbital eccentricities but the same periapse. This shared periapse is the ``collision point'' where
the impact originally occurred. 
For many orbital periods after the impact, dust continues to be generated at the collision
point, as fragments repeatedly
return there and grind down to smaller sizes (\citealt{jackson14}; \citealt{kral15}). ``The gift that keeps on giving,'' the collision point acts as a delta-function source of dust
as long as it is not smeared out by differential precession of fragment orbits.

Dust from a giant impact is more strongly apsidally aligned than dust from a parent body ring made eccentric by a planet. In the latter case, dust is created from collisions occurring across the ring over a wider (order-unity) range of orbital azimuths, and this dispersion of periapse directions is further amplified by radiation pressure and/or stellar wind drag. Accordingly, the assumption made by Lee \& Chiang (\citeyear{lee16}; see also \citealt{esposito16}) in their subset of ``periapse only'' models---that dust is created at just one orbital phase of a planet-perturbed eccentric ring---is unrealistic. Abandoning this assumption prevents those disk morphologies that hinge on strict alignment of dust grains---in particular their ``double wing''---from being realized. Double-winged disks include HD 32297 and HD 61005 (a.k.a.~``The Moth''; see \citealt{schneider14} for an atlas of debris disk images, and references therein). \citet{lin19} identified this problem and proposed that dust apsides could be re-aligned by drag from gas created in-situ by collisions in an eccentric parent body ring, and assumed to occupy apse-aligned streamlines. However, as these authors point out, gas-enforced alignment has its own problems, among them momentum feedback of dust on gas, and the fact that no gas has been detected for HD 61005 despite pointed searches \citep{olofsson16,riviere16,macgregor18}. Thus double-winged disks like HD 32297 and HD 61005 no longer seem good candidates for gravitational perturbations by a planet.

By contrast, the apsidal alignment from a giant impact is practically automatic, a consequence of the common periapsis at the collision point. A case for a giant impact has been made for the $\beta$ Pic debris disk. In scattered optical light on $\sim$1000 au scales, the system exhibits a needle whose long end points northeast (\citealt{larwood01}; \citealt{janson21}), consistent with the collision point, i.e. the delta-function dust source, being located to the southwest. Indeed an overdensity in CO, C I, and the sub-mm dust continuum is observed $\sim$80 au from the star toward the southwest  \citep{dent14,cataldi18}, in addition to a clump in the mid-infrared at $\sim$52 au in the same direction \citep{telesco05,li12}. Another disk that may have been impacted by a relatively recent catastrophic collision is AU Mic; here a collision point is hypothesized to seed dust avalanches triggered by the host M dwarf's wind 
\citep{chiang17}. Giant impacts may also explain the spatially unresolved, time-variable infrared excesses in ID8 \citep{meng14} and V488 Persei \citep{rieke21}. Detecting catastrophic collisions occurring $<10$ au from the star was anticipated from simulations of terrestrial planet formation (\citealt{kenyon05}; but see also \citealt{najita23} for ways to erase the signature of warm dusty debris).


Our goal in this paper is to survey how the debris from a giant impact appears in scattered light. We synthesize scattered light images of dust created from a singular collision point, and also from a collisional family whose member orbits have been perturbed nodally and apsidally (similar to how asteroid collision families have been perturbed by solar system planets to create the zodiacal dust bands; \citealt{dermott84}). We compare our model images with Hubble Space Telescope (HST) images of the double-winged systems HD 32297 and HD 61005, and also assess whether giant impacts can explain the needle-shaped and by implication apse-aligned systems HD 106906, HD 15115 (``The Blue Needle''), and HD 111520. Section \ref{sec:model} describes how we compute scattered light images of the debris from giant impacts, and surveys disk shape as a function of viewing geometry. We also present some thermal emission maps, suitable for comparison with the James Webb Space Telescope (JWST) and the Atacama Large Millimeter Array (ALMA), of the dust generated from collision points. Section \ref{sec:compare} compares our model images against HST and Gemini Planet Imager data for our five giant impact candidate disks. Section \ref{sec:hidden} explores what giant impacts on $\sim$100 au scales implies for distant unseen mass reservoirs. Section \ref{sec:sum} summarizes.

\section{Model}
\label{sec:model}

We consider two scenarios for the dust created in the aftermath of a giant impact. In the first (\S\ref{subsec:procedure}), dust is created from larger collision fragments whose orbits have not differentially precessed, either nodally or apsidally. Such a situation describes a giant impact that occurs away from any perturbing mass (a planet or a disk) that would induce differential precession; or, if such a perturber is present, the impact debris is not yet old enough to have differentially precessed. In the second set of models (\S\ref{sec:precess}), we consider dust generated from collisional fragments whose orbits have been differentially precessed by an inclined and possibly eccentric perturber. Some of our model assumptions are reviewed in \S\ref{subsec:model_review}.

A shortcoming of our model is that our model images are not ``flux-calibrated'': surface brightnesses are plotted in code units and not in physical units (e.g.~erg/s/cm$^2$/sr). We focus instead on relative brightnesses within an image, i.e. disk shapes. Relatedly, our models do not specify the mass released as fragments and dust in a giant impact. For our model images to apply to observations, the catastrophic disruption of a body must produce enough dust to alter a system's scattered-light appearance at optical wavelengths, beyond what a 
``background'' disk of parent bodies and dust contributes. This is an assumption that needs to be quantified and tested against observation in future work. For now we note that ``giant impacts'' might only involve ``dwarf planets'' and still produce observable scattered-light features. For example, all of the H-band light in HD 32297 may arise from a dust mass of just $\sim$$0.003 M_\oplus$, comparable to a Pluto mass, if the density of dust grains is 1 g/cc \citep{duchene20}.  

\subsection{An Unprecessed Pinched Ring and its Singular Collision Point}\label{subsec:procedure}

Our starting point is the ``pinched ring''
of debris produced in the immediate aftermath of a catastrophic
collision (\citealt{jackson14}, their figure 8). The ring is composed
of collision fragments large enough to be unaffected by stellar radiation pressure, and that have differentially rotated (but not differentially precessed) around the star because of
small differences in their semimajor axes. The ring is thus at
least a few dozen orbital periods old, but not so old that the
fragment orbits have differentially precessed (which they eventually would if in the presence of a nearby perturbing mass; see \S\ref{sec:precess}). The ring is narrowest
(pinched) at the location of the originating impact---this is the
``collision point'' from which fragments are initially launched, and
to which they repeatedly return on closed orbits. The collision point,
which does not move in inertial space, represents a quasi-singularity in the ring where
fragment densities and velocity dispersions, and therefore collision rates, are highest.\footnote{Azimuthally opposite the collision point there is an ``anti-collision line'' where parent body orbits converge vertically on the mean orbital plane;  parent body densities here are also relatively high, but not as high as at the collision point where there is radial in addition to vertical convergence. Note further that dust grains (not parent bodies) on orbits made elliptical by stellar radiation pressure and wind drag converge on the collision point but not the anti-collision line.}

Our
model assumes fragment collisions and by extension dust production are dominated
by the collision point,
and ignores collisions elsewhere
along the ring (this assumption will be relaxed in \S\ref{subsubsec:proc_fam} when we treat precessed collisional families). We now describe how to compute the orbits of dust grains and synthesize disk images in both in scattered light and thermal emission.

\subsubsection{Procedure for Generating Scattered Light Images of Dust from a Collision Point Singularity} \label{subsubsec:proc}
Step 1 of our procedure 
is to compute a set of $N_{\rm p} = 200$ orbits (sometimes higher for greater resolution) of ``parent bodies''---large collision fragments comprising the pinched ring, unaffected by stellar radiation pressure or stellar wind drag---launched from
the single collision point at radius $\mathbf{R}$ from the star. The parent bodies 
have initial velocities $\mathbf{V}$ equal to a shared base velocity
$\mathbf{V}_0$, plus a random
isotropic component $\mathbf{\delta V}$ that varies from body to body. The base velocity is the orbital velocity of
the progenitor just before its destruction,
and the random component accounts for the dispersion of ejecta velocities from the original impact. 
We fix $|\mathbf{\delta V}| = 0.025
|\mathbf{V_0}|$ (implying that eccentricities of parent bodies range from 0 to 0.025).  Our fiducial model assumes
a circular progenitor orbit around a solar-mass star at $|\mathbf{R}| = 60$ au, implying $|\mathbf{V}_0| \simeq 4$ km/s and $|\mathbf{\delta V}| \simeq 100$ m/s; the last quantity is 
of order the 
ejecta velocities inferred for the Haumea collisional family in the Kuiper belt \citep{lykawka12}. Our model assumptions are made for simplicity
 and subject to revision, as ejecta velocities $\mathbf{\delta V}$ may
 not be isotropic \citep{watt21} and may need to be higher to
 ensure that fragments can escape the gravity well of the
 progenitor. Insofar as it concerns the scattered light images, these two errors tend to cancel: increasing $|\mathbf{\delta V}|$ while retaining isotropy blurs scattered light images, while allowing for anisotropy sharpens them.

Step 2 is to compute from our set of $\{ \mathbf{R,V} \}$ (having $N_{\rm p}$ entries) the
corresponding set of Kepler orbital elements  $\{ a, e, i, \Omega,
\omega, f \}$ using the gravitational point-mass potential from the
host star. Our notation here is standard \citep{murray_dermott}, as
are the equations (listed in the Appendix) needed to perform this computation.

The rest of our procedure is similar to that of \citet{lee16}, but with an improved scheme for sampling the radiation+wind-to-gravity force ratio $\beta$ for dust grains. In step 3, for each of our $N_{\rm p}$ parent bodies located at the collision point $\mathbf{R}$, we compute the orbital elements $\{ a', e', i',
\Omega', \omega' \}$ of $N_\beta = 5000$ ``dust orbits'' launched from $\mathbf{R}$ --- these share the same initial $\{ \mathbf{R,V} \}$ as the parent bodies, but feel a gravitational force reduced by $(1-\beta)$ (see the Appendix for explicit expressions of primed quantities). 
The true anomalies $\{f'\}$ of the individual dust grains populating these orbits are specified later in step 4.
The force ratio $\beta$ depends on grain size $s$; we assume that grains absorb and scatter stellar photons and/or stellar wind particles geometrically, so that $\beta \propto 1/s$ (this follows from the radiation+wind pressure force scaling as grain area $s^2$ and gravity scaling as grain mass or volume $s^3$ at fixed internal grain density; see  \S\ref{subsec:model_review} for further discussion). We adopt the modified Dohnanyi size distribution of \citet{strubbe06}, appropriate for ``halo grains'' whose orbits are highly elongated by radiation and/or stellar wind pressure. Halo grains are relatively immune to destructive collisions as they traverse the rarefied regions near their apastra, and the consequent prolonging of their lifetimes enhances their steady-state population. Whereas a standard Dohnanyi differential size distribution predicts $dN/d\beta \propto \beta^{3/2}$ (this is equivalent to $dN/ds \propto s^{-7/2}$), grains whose orbits extend from the collision point into the distant halo have their numbers enhanced in proportion to their orbital period:
\begin{eqnarray}
dN/d\beta & \propto & \beta^{3/2} \times (a')^{3/2} \nonumber \\
& \propto & \beta^{3/2} \frac{(1-\beta)^{3/2}}{ \left[ 1 - e^2 - 2\beta (1+ e \cos f) \right]^{3/2}} \label{eq:orbcorr}
\end{eqnarray}
(see also \citealt{lee16}, and our Appendix). This $\beta$-distribution formally cuts
off at 
\begin{equation}
\beta_{\infty} = \frac{1-e^2}{2(1+e \cos f)} \,,
\label{eq:betamax}
\end{equation}
the value corresponding to a marginally
bound (zero energy; $e'=1$) orbit of infinite period, and for
which $dN/d\beta$ is infinite. In reality the distribution
cuts off at a value $\beta_{\rm max} < \beta_{\infty}$ determined by
the finite 
age of the system and physical conditions at large distance (e.g.~the
influence of the interstellar medium, or the Galactic tide). Unless otherwise specified we set
$\beta_{\rm max} = 0.997\beta_{\infty}$, implying for our fiducial parameters 
a maximum apastron distance for a dust grain orbit on the order of 10000 au. In \S\ref{subsec:model_review}, we review the arguments supporting our use of equations (\ref{eq:orbcorr})--(\ref{eq:betamax}).

Thus in step 3 of our procedure, we compute, for each of our $N_{\rm p}$ parent bodies, a different $\beta_{\rm max}$. For each such $\beta_{\rm max}$, we then randomly draw
$N_\beta$ values of $\beta$ from the distribution (\ref{eq:orbcorr}) evaluated between $\beta_{\rm min} = 0.001$ and $\beta_{\rm max}$. 
Each draw requires us to interpolate the cumulative distribution
obtained by integrating $dN/d\beta$; this cumulative distribution is
evaluated on a logarithmic grid in $\beta$ which concentrates its 
precision near $\beta_{\rm max}$ where $dN/d\beta$ is steepest (the
distribution is very top-heavy; see also \citealt{olofsson22}). The values of $\beta$ so drawn enable us to convert
$\{a,e,i,\Omega,\omega,f\}$ to $\{ a',e',i',\Omega',\omega'\}$ (see the Appendix).

Step 4 populates each of our $N_{\rm p} \times N_{\beta}$ dust orbits with $N_{\rm dust-per-orbit} = 100$ dust grains. We randomly draw their mean anomalies $\{M'\}$ from a uniform distribution between 0 and $2\pi$, and convert to $\{f'\}$ using Kepler's equation.

Step 5 takes our $N_{\rm dust} = N_{\rm p} \times N_\beta \times N_{\rm dust-per-orbit}$ dust particles and projects them onto the sky plane of a distant
observer to synthesize a scattered light image.
The sky plane of 800 $\times$ 800 AU, centered on the star,
is divided into 800 $\times$ 800 square cells, and each dust particle
contributes, to the cell in which it is located,
a surface brightness (units of energy per area per time per solid angle)
proportional to $\phi(\theta)/(\beta^2 r^2)$. 
Here $1/\beta^2$ accounts for the scattering cross
section for each grain (assumed geometric),
$r$ is the distance between the dust particle
and the host star, and $\phi(\theta)$ is the 
\citet{hedman15} scattering
phase function (SPF), where $\theta$ equals 
the scattering angle between the vector joining the star to the dust particle and the line-of-sight vector to the observer. This SPF was derived in the optical continuum using
Saturn's irregularly shaped, 
strongly forward-scattering G ring particles, and approximates the SPFs inferred for many debris disks \citep{hughes18}, including HD 61005 \citep{esposito16}.

Occasionally, for the purpose of comparing with GPI (Gemini Planet Imager) data, we will have need of an infrared SPF, which in H-band is expected to be more isotropic than in the visible (insofar as diffracted beam widths scale linearly with wavelength). We mock up such a function by arbitrarily adjusting the coefficients of the Hedman-Stark SPF, reducing the weight $w_1$ of their strongly forward-scattering component to 0.513 (from the original 0.643; see their Table 7), and increasing the weight $w_3$ of their nearly isotropic component to 0.311 (from the original 0.181). Compared to the original visible-wavelength Hedman-Stark SPF which beams 50\% of its power into scattering angles $\theta < 1.24^\circ$ and 75\% into $\theta < 34.1^\circ$, our mock infrared SPF beams those relative powers into wider angles $\theta < 6.28^\circ$ and $\theta < 64.9^\circ$, respectively. The overall shape of this infrared SPF is otherwise similar to the visible SPF (cf.~\citealt{chen20}).

The orientation of the observer on the celestial sphere centered on the host star is parameterized by altitude Alt (the inclination angle relative to the debris field’s mean orbital plane) and azimuth Az (angle measured in the debris mean plane). Alt = $0^\circ$ corresponds to the disk seen edge-on, and Alt = $\pm$90$^\circ$ give face-on views. Az = 0$^\circ$ corresponds to the line joining the star to the collision point pointing away from the observer, while Az = 180$^\circ$ directs the collision point toward the observer.

\subsubsection{Fiducial Results for Scattered Light Images of Dust from a Collision Point Singularity}\label{subsec:fiducial}

\begin{figure*}[ht!]
\plotone{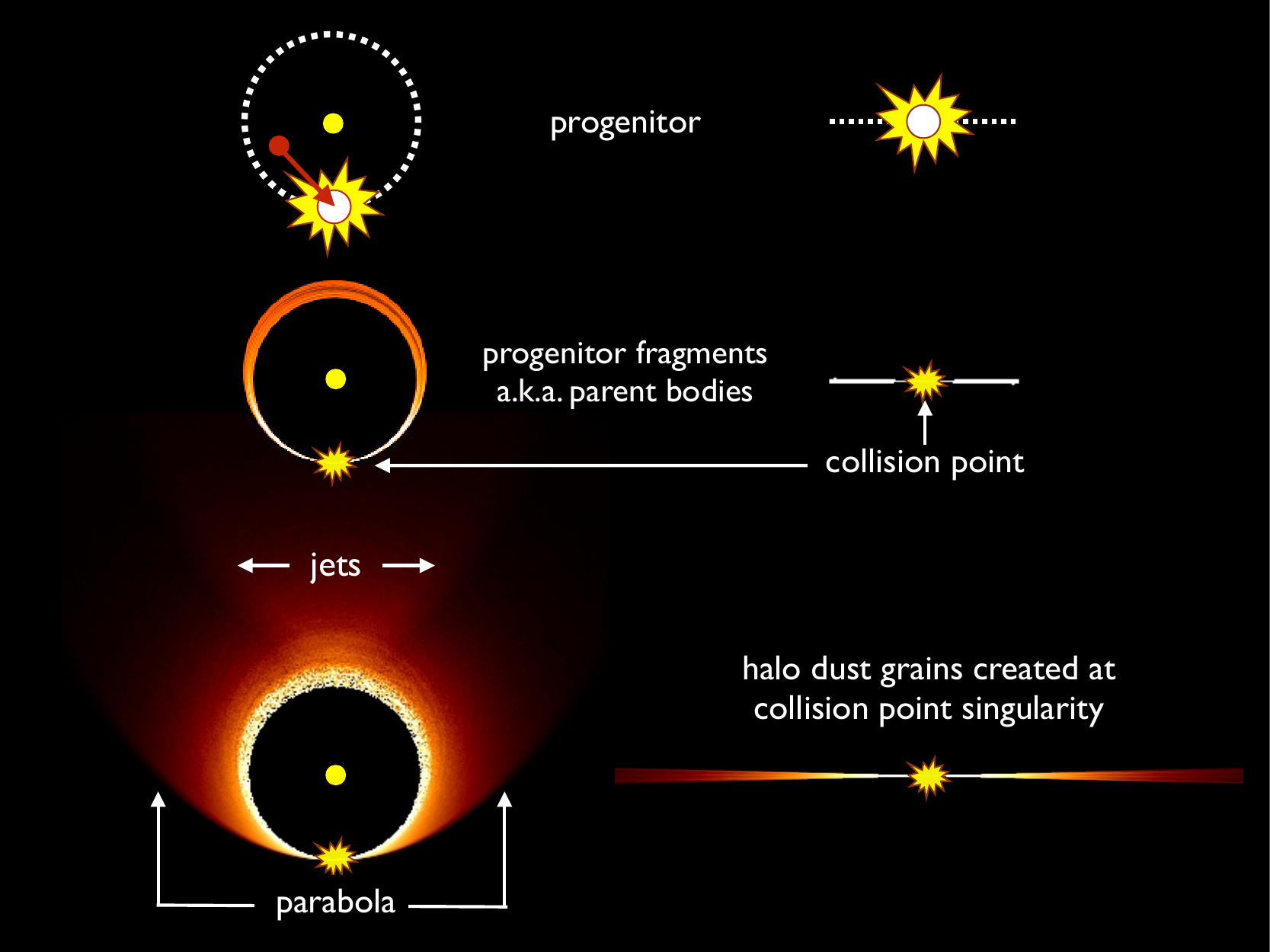}
\caption{Aftermath of a giant impact, shown schematically. The destruction of a progenitor body (e.g.~a dwarf planet) produces fragments, termed ``parent bodies'' in our light scattering code, that differentially rotate into a pinched ring. Face-on views are shown in the left column, and edge-on views in the right. The ring is pinched at the original collision point where fragments were launched and continually return on closed Keplerian orbits. The scattered light images at the bottom of the figure are of dust created at the collision point, assuming there are no sources of differential precession that smear away the collision point singularity (see \S\ref{sec:precess} and Figures \ref{fig:precessed_eccentric} and \ref{fig:precessed_circular} where we relax this assumption). 
\label{fig:aftermath_pdf}}
\end{figure*}

Figure \ref{fig:aftermath_pdf} shows face-on and edge-on views of the
parent body fragments (middle row) and dust grains (bottom row)
resulting from the catastrophic disruption of a progenitor whose orbit was circular. We see the pinched ring
occupied by the parents, and the convergence of both parent body and
dust grain orbits on the singular collision point.

The dust grain scattered light maps in Fig.~\ref{fig:aftermath_pdf}
are generated using
$N_{\rm dust} = N_{\rm p} \times N_\beta \times N_{\rm dust-per-orbit} = 200 \times 5000 \times 100 = 10^8$
particles. The parabola forming the outer boundary of the
dust map is traced by the smallest particles that are still bound to the
star and which feel the strongest radiation+wind pressure, having
$\beta \simeq \beta_{\rm max}$.  We also see twin ``jets'' emerging
from the backside. Discussed by \citet{lee16} in their
``periapse only'' models (see their figure 7), the jets are particle
overdensities resulting from two competing effects: the tendency for
grains to maximize their density at apastron because they spend more
time there, and the opposing tendency to concentrate at periastron,
near the collision point where all orbits cross. The result is that
particle densities maximize at neither apastron nor periastron, but at
two intermediate orbital angles, reflection symmetric about the common
apsidal line.

\begin{figure*}[ht!]
\plotone{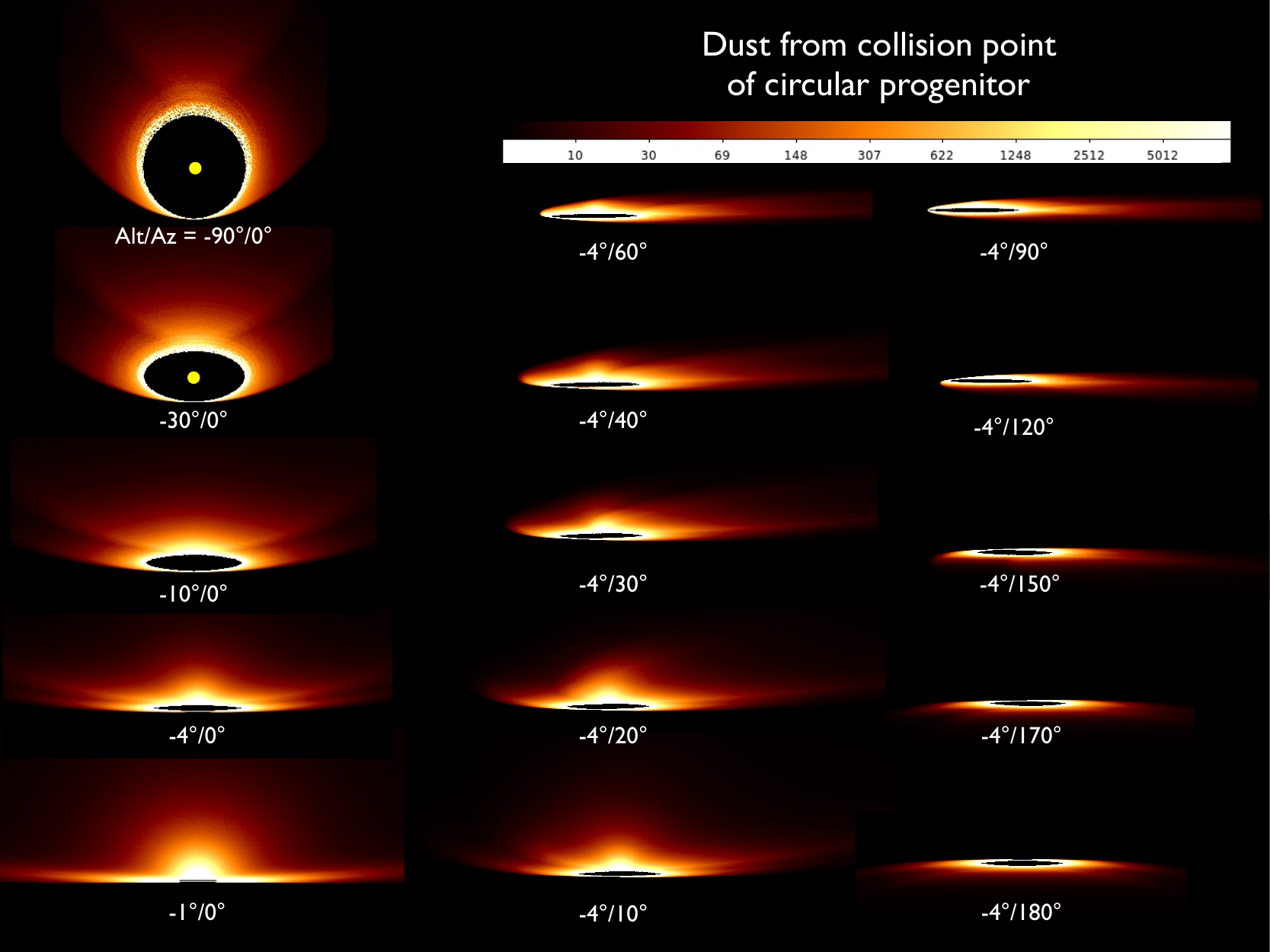}
\caption{Scattered light images of dust grains created from a collision point singularity, assuming a circular orbit for the progenitor. The images derive from $N_{\rm p} = 200$ parent bodies, $N_\beta = 5000$ dust orbits per parent body, and $N_{\rm dust-per-orbit} = 100$ dust grains per dust orbit, for a total of $N_{\rm dust} = 10^8$ dust particles. Surface brightness is plotted on a log scale that saturates (does not resolve) the brightest regions near the collision point, but otherwise resolves about three orders of magnitude in brightness (see color scale bar). From top to bottom in the left column, the disk tilts to point the jets toward the observer, while the pinched collision point tilts away from the observer. The parabola and jets seen in a face-on view (Az = 0, Alt $\sim -90^\circ$) are projected into a set of double wings in nearly but not exactly edge-on views (Az = 0, $|{\rm Alt}| \lesssim 4^\circ$).  Varying Az in the middle and right columns generates left-right asymmetries;  the needles seen at Az $\sim$ 40$^\circ$--120$^\circ$ arise from the jets extending to the right of the star, and the collision point lying to the left. Halo grains are brighter when viewed in forward-scattered (Az $\sim 0^\circ$) than in back-scattered (Az $\sim 180^\circ$) light. Only Alt $<0$ is shown here, which places the observer below the dust orbital plane; identical images flipped top to bottom result if Alt $>0$. Animation of this figure available at  \url{https://github.com/joshwajones/jones_etal_animations/}.
\label{fig:pano}}
\end{figure*}

Figure \ref{fig:pano} displays dust grain scattered light maps at various viewing
orientations. At fixed Az = 0 (leftmost column of images), as Alt changes from -90$^\circ$ toward a more edge-on view at -1$^\circ$, the halo grains are directed increasingly toward the observer, in front of the star. Note that Alt $<0$ corresponds to the observer situated below the dust orbital plane; identical images flipped top to bottom result for Alt $>0$.  Forward scattering of halo grains creates a ``bulb'' above the star at $-4^\circ \lesssim {\rm Alt} \lesssim -1^\circ$ (not shown is the view for Alt = $0$ when the bulb disappears because the dust occupies too thin a sheet). A ``double wing'' also appears for nearly edge-on views: the parabola projects into a ``primary wing'' with a curved spine below the star, and the jets project into a ``secondary wing'' running nearly parallel to the primary wing, above the star. The two wings merge for Alt between $-1^\circ$ and 0. Taking our nearly edge-on view at fixed Alt = $-4^\circ$, and turning it in Az from 0 to 180$^\circ$ generates left-right asymmetries (middle and right columns in Fig.~\ref{fig:pano}), maximized for Az = $90^\circ$. As Az increases from $90^\circ$ to 180$^\circ$, the halo grains are seen increasingly in back-scattered light and so appear dimmer.

\begin{figure*}[ht!]
\plotone{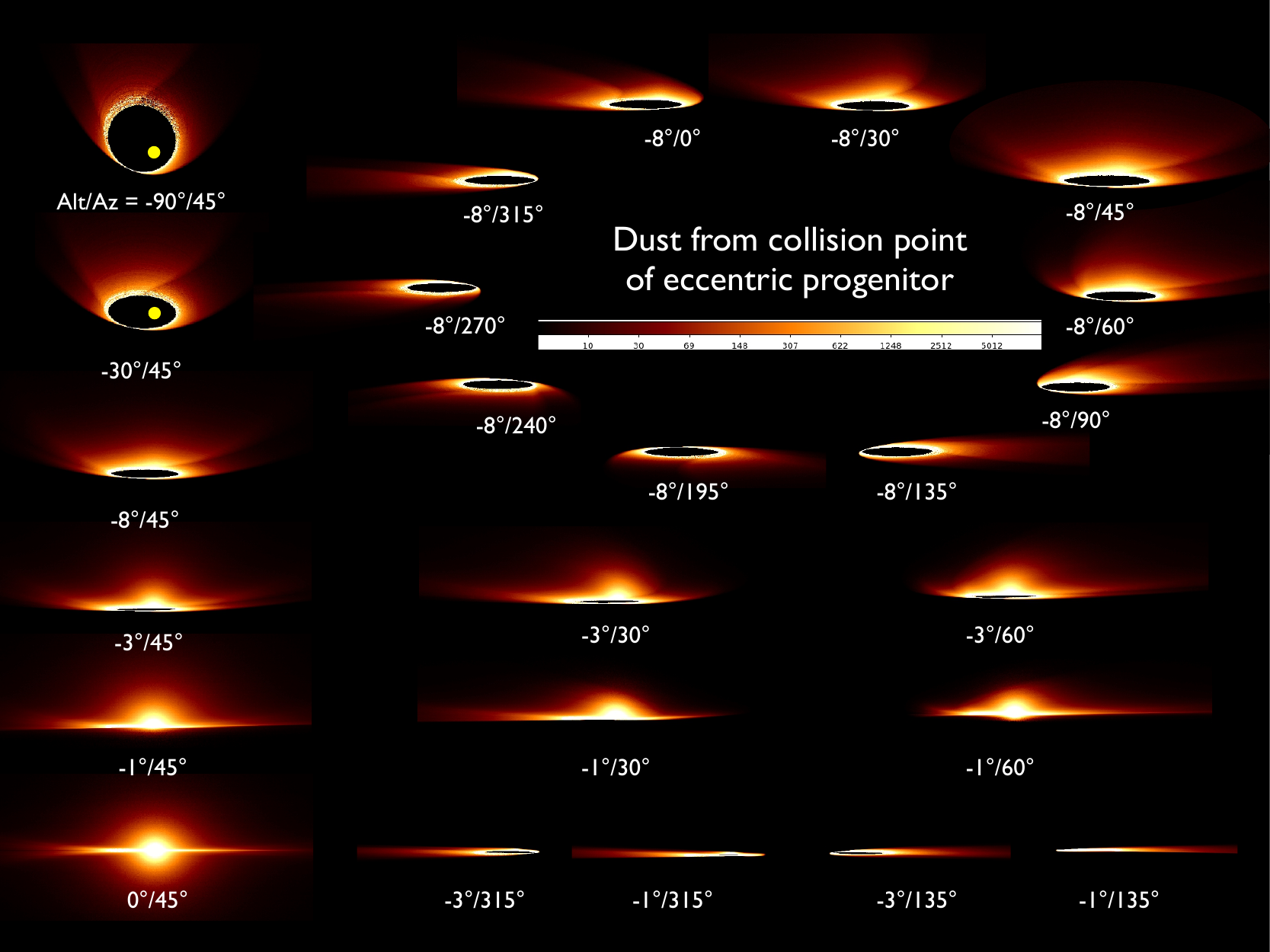}
\caption{Same as Fig.~\ref{fig:pano}, but for a progenitor with eccentricity = 0.4, shattered when its true anomaly was $\pi/2$ (quadrature); see the face-on view at top left showing the collision point where the ring is pinched. In this view, the direction of orbital motion is clockwise. Viewing Alt is varied over the leftmost vertical column of images which tilt the ring so that the collision point is behind the star and most of the dust is in front of the star; Alt $<0$ so that the observer is looking from below the mean dust plane. At fixed Alt = -90$^\circ$, increasing Az turns the disk clockwise. The circle of images to the top right varies the viewing azimuth Az  at fixed Alt. At Az = 0, the pinched collision point is behind the star and to the right; at Az $\sim 30^\circ$, it is almost directly behind the star; at Az $\sim 90^\circ$, it lies to the left; and at Az $\sim 195^\circ$, it is nearly directly in front of the star. The dust halo is brighter in forward-scattered light when the collision point is located behind the star (Az $\sim 30-45^\circ$) than when the collision point is in front (Az $\sim 195-240^\circ$). There are left-right asymmetries at all viewing orientations, a consequence of progenitor eccentricity which distorts the parabola and jets seen for a circular progenitor (cf.~Fig.~\ref{fig:pano}). Note, e.g., how the left ansae are vertically thicker and radially shorter than the right ansae in the images at bottom left for Az = 45$^\circ$ and Alt = 0, -1$^\circ$. We have verified that if the progenitor is shattered when its true anomaly was $-\pi/2$, essentially the same morphologies shown in this figure are generated, just rotated in azimuth. 
\label{fig:pano_ecc}}
\end{figure*}

The images in Figs.~\ref{fig:aftermath_pdf} and \ref{fig:pano} derive from a progenitor on a circular orbit (the progenitor orbit determines the magnitude and direction of $\mathbf{V_0}$ in step 1 of our procedure). Figure \ref{fig:pano_ecc}
 displays images generated using a progenitor having an eccentricity of 0.4, collisionally disrupted when its true anomaly was $\pi/2$ (quadrature; we have verified that we obtain essentially identical results for a true anomaly of $-\pi/2$ by reflection symmetry). For the eccentric case, we see the same features as in the circular case---a bulb and double wings---but distorted so that left-right asymmetries are evident for all views.

\subsubsection{Thermal Emission Images of Dust \\from a Collision Point Singularity}\label{subsubsec:thermal}

Our procedure for generating scattered light images is readily modified to produce thermal emission images at wavelength $\lambda$. None of steps 1--4 for laying down dust grains in space changes. Only in step 5 do we replace the scattered light contribution per particle of $\phi(\theta)/(\beta^2 r^2)$ with its thermal emission counterpart:
\begin{align}
\frac{\phi(\theta)}{\beta^2 r^2} \rightarrow &
\, \frac{\varepsilon}{\beta^2 \{ \exp[hc/(\lambda k T)]-1 \} } 
\nonumber 
\label{eqn:replace}
\end{align}
where the factor of $\beta^2$ arises as before from the geometric cross section of a dust particle, $\varepsilon$ is the dust emissivity, and the term in braces comes from the Planck blackbody function for dust temperature $T$, Boltzmann constant $k$, Planck's constant $h$, and speed of light $c$. The resultant thermal emission images are in code units for specific intensity (energy per area per time per solid angle per frequency) as distinct from our scattered light images which are in code units for surface brightness (energy per area per time per solid angle).

The emissivity $\varepsilon$ is modeled as a simple function of $\lambda$ and dust particle radius $s$:
\begin{equation}
\varepsilon (s, \lambda) = \begin{cases}
1 & {\rm if} \,s > \lambda/(2\pi) \\
2\pi s/\lambda & {\rm if} \,s \leq \lambda/(2\pi) \nonumber
\end{cases}
\end{equation}
where $s$ and $\beta$ are related by 
\begin{equation}
s = \frac{3}{16\pi} \frac{L_\star}{GM_\star c\rho \beta} \nonumber
\end{equation}
and $G$ is the gravitational constant. 
Because the broken power-law emissivity introduces a scale ($\lambda$) into the problem, we need to be definite about parameters: for our fiducial model we assign $L_\star = 6.6 L_\odot$ and $M_\star = 2.7 M_\odot$ for the host stellar luminosity and mass (as appropriate for HD 106906), and take the grain internal density $\rho$ to be 1 g/cm$^3$ (this is lower than for silicates with $\sim$3 g/cm$^3$, but may still be reasonable given lower-density ices and/or grain porosity).

The dust temperature is given by:
\begin{equation}
T = \left( \frac{L_\star}{16\pi \sigma r^2} \right)^{1/4} \left( \frac{1}{\langle\varepsilon\rangle_T} \right)^{1/4} \nonumber
\end{equation}
where $\sigma$ is the Stefan-Boltzmann constant and we approximate the Planck-averaged emissivity 
as $\langle \varepsilon \rangle_T = \min (2\pi s/[hc/(3kT)], 1)$, using $hc/(3kT)$ as the wavelength where the Planck function $B_{\nu}$ peaks.

\begin{figure*}[ht!]
\plotone{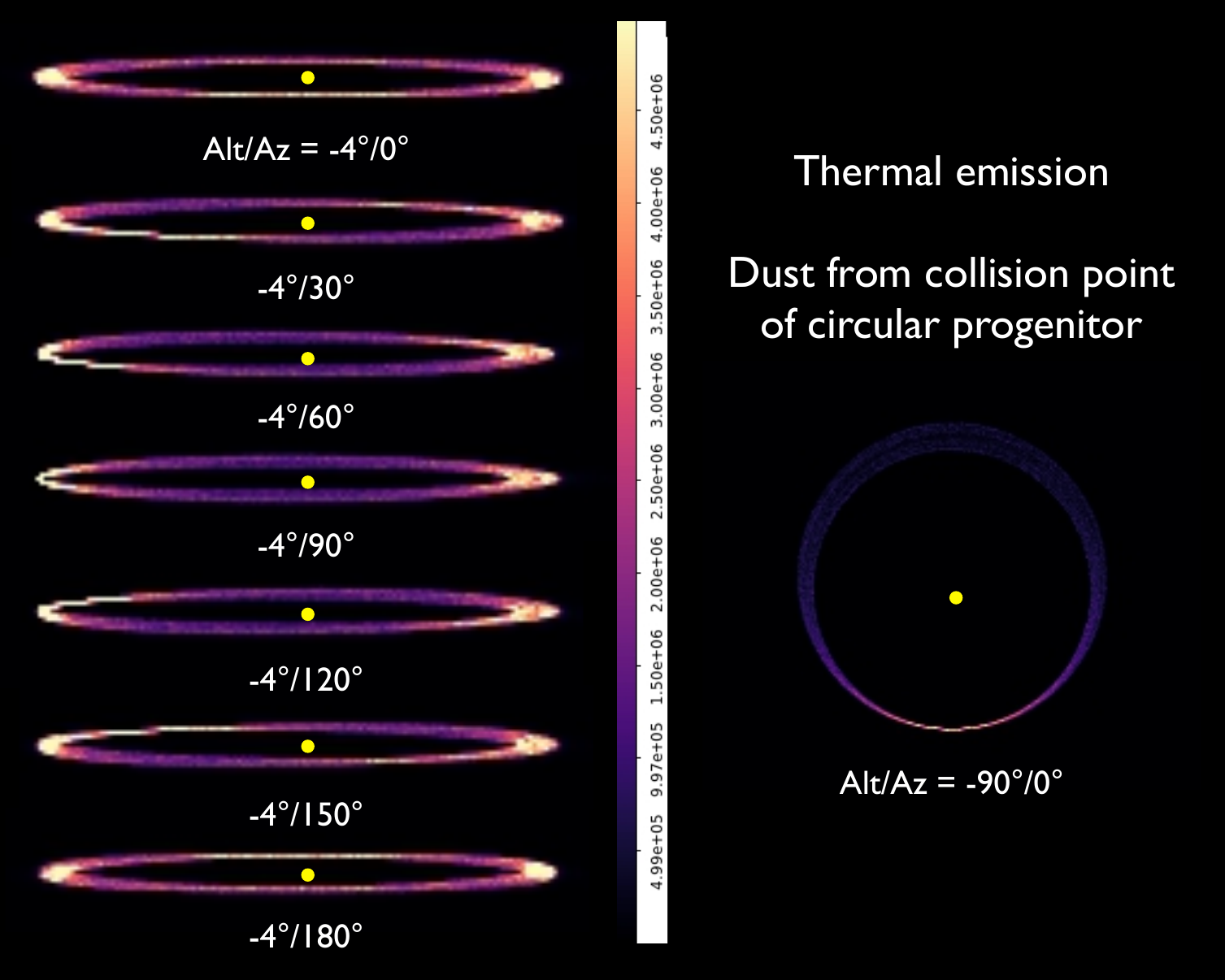}
\caption{Thermal emission from dust grains created from a collision point singularity, assuming a circular orbit for the progenitor. The color scale is linear and in code units for specific intensity (energy per time per area per solid angle per frequency), resolving about a factor of 10 from minimum to maximum intensity.  Images are generated for $\lambda = 1$ mm but look practically identical for $\lambda = 25$ $\mu$m (aside from a change in brightness normalization) because the emission is dominated by the hottest grains lying nearly equidistant to the star at nearly uniform temperature; thus variations in surface brightness are due mostly to local variations in line-of-sight column density. The region near the collision point (projected below the star at Az = 0; to the extreme left at Az = 90$^\circ$; and above the star at Az = 180$^\circ$) is brighter than its surroundings by factors of a few.
\label{fig:thermal}}
\end{figure*}

Figure \ref{fig:thermal} displays thermal emission images generated for our fiducial model of a circular progenitor at $\lambda = 1$ mm. Images generated at $\lambda = 25$ $\mu$m (not shown) are practically identical aside from a change in flux normalization, because the emission is dominated by the hottest dust closest to the star at nearly constant $r$, and variations in brightness are due primarily to variations in line-of-sight column density, not temperature. For the nearly edge-on disk as shown, those column densities are highest at the limbs and the collision point. The collision point is brighter than its surroundings by factors of a few. As Az varies from 0 to 180$^\circ$, the overbright collision point turns clockwise from below the star to above the star.

\vspace{0.3in}
\subsection{Precessed Collisional Family}\label{sec:precess}

If the pinched ring is near a massive planet or disk, the orbits of the fragments making up the ring are forced to differentially precess, nodally and apsidally, smearing away the collision point singularity. The fragments comprising the collisional family then have orbital elements that can be decomposed into forced terms (forced by the perturbing mass) and free terms determined by initial conditions (e.g.~\citealt{murray_dermott}, section 7.10). Family members are identified by their shared semi-major axes, free eccentricities, and free inclinations. In the solar system, the IRAS (Infrared Astronomical Satellite) zodiacal dust bands are attributed to the differentially precessed Koronis, Eos, and Themis asteroid families \citep{dermott84,sykes86,nesvorny03}.

\subsubsection{Procedure for Generating Scattered Light Images of Dust from Precessed Collisional Families}\label{subsubsec:proc_fam}

Steps 1 and 2 of the procedure described in \S\ref{subsubsec:proc} for unprecessed debris are replaced as follows for precessed debris. The
eccentricities $\{e\}$ and longitudes of periastron $\{\varpi\}$ of $N_{\rm p}$ parent bodies are given by
\begin{eqnarray}
e \sin \varpi &= e_{\rm forced} \sin \varpi_{\rm forced} + e_{\rm free} \sin \varpi_{\rm free}\\
 e \cos \varpi &= e_{\rm forced} \cos \varpi_{\rm forced} + e_{\rm free} \cos \varpi_{\rm free}
\end{eqnarray}
with $e_{\rm forced}$,
$\varpi_{\rm forced}$, and $e_{\rm free}$ identical for all parent bodies, and $\varpi_{\rm free}$ drawn randomly from a uniform distribution between 0 and $2\pi$ \citep{murray_dermott}. The uniform distribution of $\varpi_{\rm free}$ describes fragments that have had their apsidal longitudes (technically their free longitudes, not their osculating longitudes) differentially precessed. Inclinations 
$\{i\}$ and longitudes of ascending node $\{\Omega\}$ are specified analogously:
\begin{eqnarray}
 i \sin \Omega &= i_{\rm forced} \sin \Omega_{\rm forced} + i_{\rm free} \sin \Omega_{\rm free}\\
 i \cos \Omega &= i_{\rm forced} \cos \Omega_{\rm forced} + i_{\rm free} \cos \Omega_{\rm free}
\end{eqnarray}
with $\Omega_{\rm free}$ uniformly distributed from 0 to $2\pi$.

We consider two cases:  $\{ e_{\rm forced} = 0.3$, $\varpi_{\rm forced} = 0$, $i_{\rm forced} = 0$, $e_{\rm free} = i_{\rm free} = 0.05, a = 86 \,{\rm au}\}$ which models parent bodies forced by an eccentric perturber, and $\{ e_{\rm forced} = i_{\rm forced} = 0$, $e_{\rm free} = i_{\rm free} = 0.05, a = 60 \, {\rm au} \}$ where the perturber's orbit is circular.  In the first case, parent bodies are on nearly apse-aligned, eccentric orbits with periastron distances of about 60 au, matching the scale of our other experiments. In both cases, the perturber's orbital plane defines the reference plane, with parent bodies inclined to this plane by $0.05$ rad = 2.9$^\circ$.

\begin{figure*}[ht!]
\plotone{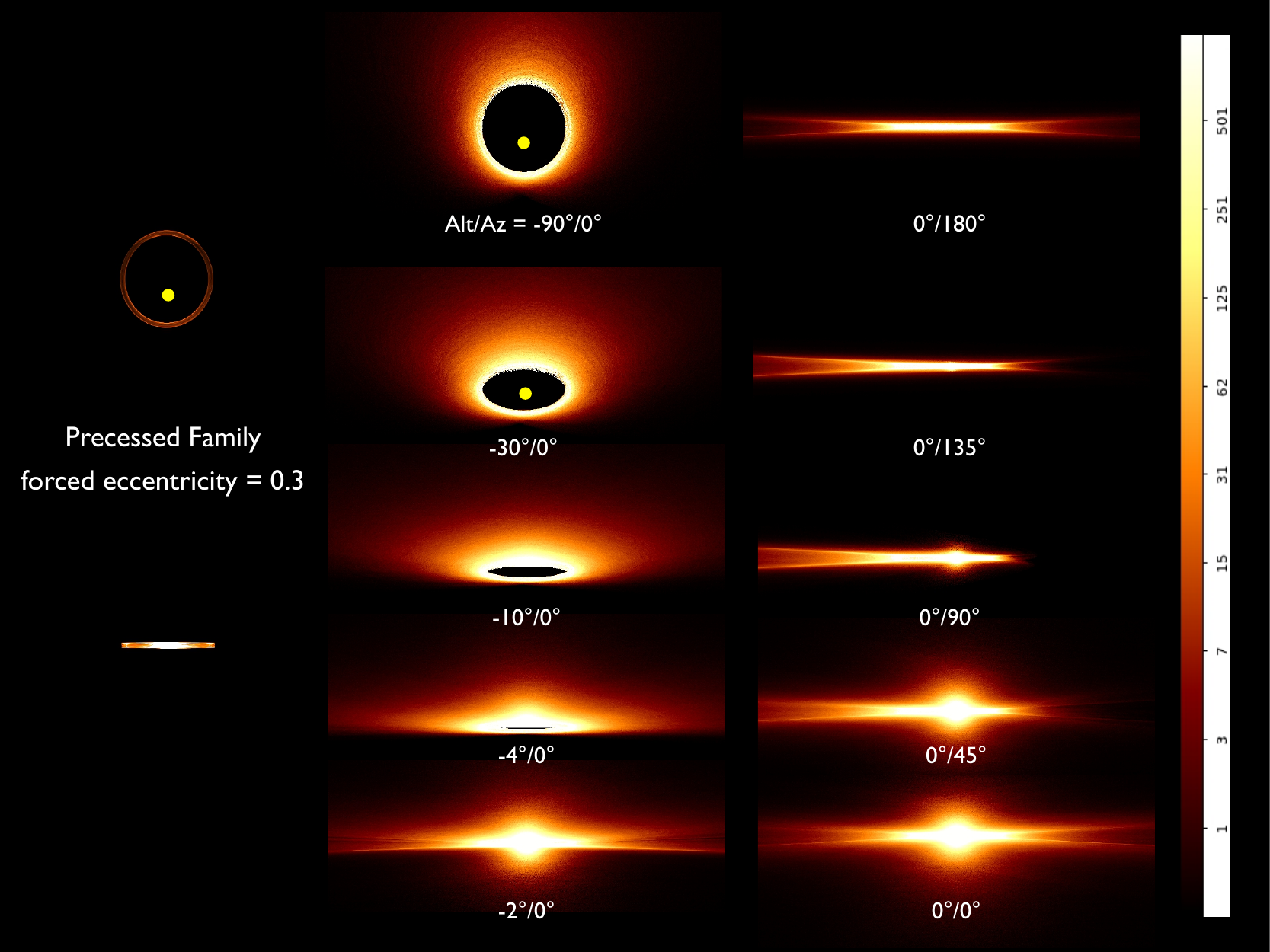}
\caption{Scattered light images from a precessed collisional family with $e_{\rm forced} = 0.3$, $i_{\rm forced} = 0$, and $e_{\rm free} = i_{\rm free} = 0.05$. The left column shows face-on and edge-on views of the parent bodies which do not feel radiation pressure or stellar wind drag (see also \citealt{wyatt99}, their figure 2 right panel, for a similar parent body picture).
Remaining images are of dust grains which do feel radiation pressure and/or wind drag, on a common log surface brightness scale 
that does not resolve (i.e.~saturates) the brightest regions.
``Forks'' are seen at low $|{\rm Alt}|$; these are the limb-brightened top and bottom surfaces of a disk of particles having fixed inclination $i$ and uniformly distributed $\Omega$. Dust particles blown by radiation pressure and/or a stellar wind form a wake in the direction of apastron; the wake can be directed toward the observer in which case there is strong forward scattering (${\rm Az} \simeq 90^\circ$); away from the observer (${\rm Az} \simeq -90^\circ$); or perpendicular to the observer's line of sight (${\rm Az} \simeq 0^\circ$ or $180^\circ$) in which case the fork is visible on only one side of the star. Animations related to this figure available at  \url{https://github.com/joshwajones/jones_etal_animations/}.
\label{fig:precessed_eccentric}}
\end{figure*}

\begin{figure*}[ht!]
\plotone{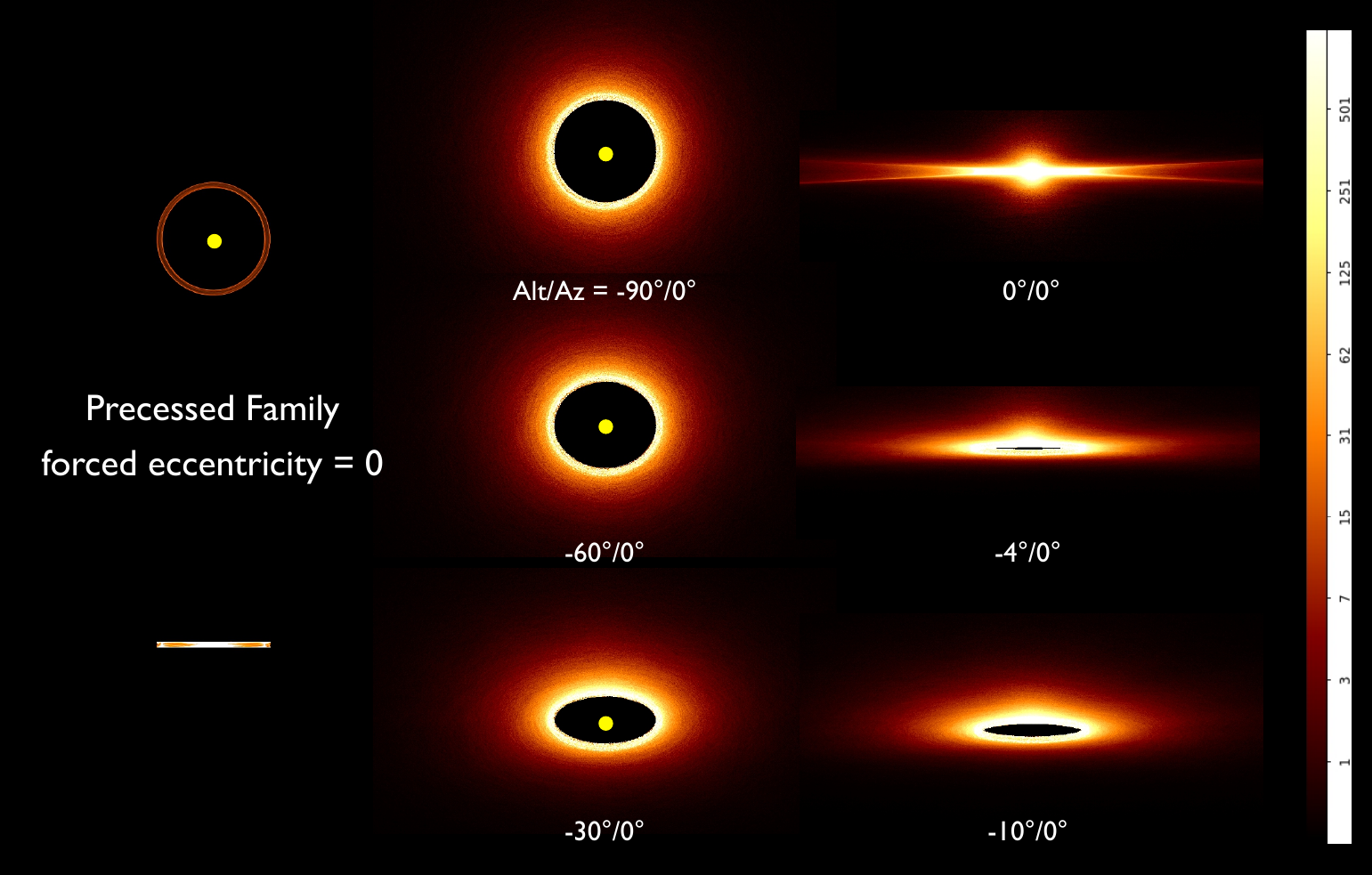}
\caption{Same as Fig.~\ref{fig:precessed_eccentric} but for a precessed collisional family with $e_{\rm forced} = i_{\rm forced} = 0$ and $e_{\rm free} = i_{\rm free} = 0.05$. The left column shows face-on and edge-on views of the parent bodies which do not feel radiation pressure or stellar wind drag.  
Remaining images are of dust grains which do feel radiation pressure and/or wind drag, on a common log surface brightness scale that saturates (does not resolve) the brightest regions.
\label{fig:precessed_circular}}
\end{figure*}

Having specified $\{a,e,i,\Omega,\omega = \varpi-\Omega\}$ of the parent bodies, we now decide $\{f\}$, the true anomalies of the parent bodies from which dust grains are launched. 
In the eccentric family case (Figure \ref{fig:precessed_eccentric}), particle densities are highest near periastra, and so we 
randomly draw $\{f\}$ from a uniform distribution between $-\pi/2$ to $\pi/2$. Our results are not sensitive to the exact limits of this distribution, as we have checked; the point is that most of the dust should be generated from a region subtending an angle on the order of $\sim$1 radian centered on the mean periastron. In the circular family case (Figure \ref{fig:precessed_circular}), the system is on average azimuthally symmetric, and so we draw $\{f\}$ randomly from 0 to $2\pi$.

The rest of our procedure follows steps 3--5 of \S\ref{subsubsec:proc}.

\subsubsection{Fiducial Results for Scattered Light Images of Dust from a Precessed Collisional Family}

Figure \ref{fig:precessed_eccentric} displays images of dust created from a precessed collisional family having a forced eccentricity $e_{\rm forced} = 0.3$; Figure \ref{fig:precessed_circular} shows the same but for $e_{\rm forced}=0$. For face-on or nearly so views, the images are similar to those from \citet{lee16}, unsurprisingly since the model ingredients are practically identical. The edge-on views differ, however; here we fix $i_{\rm free} =$ constant rather than sample $i_{\rm free}$ from a continuous distribution. A single $i_{\rm free}$ is  shared by all fragments born from a single progenitor. Fixing $i_{\rm free}$ and allowing $\Omega$ to vary produces, in edge-on or nearly so views, a disk whose top and bottom surfaces are limb-brightened---``forks'' appear on one or both ansae depending on the viewing azimuth, and whether the collisional family has a non-zero forced eccentricity. In the solar system, these forks are known as the zodiacal dust bands which come in pairs. 

\subsection{Reviewing Model Assumptions}\label{subsec:model_review}

\subsubsection{Top-heavy $\beta$-distribution}

We have used a top-heavy $\beta$-distribution in equations (\ref{eq:orbcorr})--(\ref{eq:betamax}) that assumes dust can be blown onto near-parabolic orbits by a combination of stellar radiation pressure and stellar wind pressure. Stellar radiation acting alone may not be effective at blowing grains out. Depending on grain composition, dust opacities at optical wavelengths may not be large enough for $\beta$ to approach the blow-out value in eqn.\,(\ref{eq:betamax}); see, e.g., \citet{lebreton12} and \citet{arnold19}. A parallel concern is that stellar luminosities may not be large enough for blow out, in particular in the HD 61005 system whose host star luminosity is $0.6 L_\odot$. If near-blow out is not possible, we lose the most eccentric grains having the most distant apastra, and jets and double wings would fade from images of dust created from collision point singularities.

Blow-out can still be achieved by ram pressure drag exerted by a stellar wind (e.g.~\citealt{burns79}). Grain cross sections to particle winds should be at least geometric, and possibly larger for charged grains. Young, low-mass, strongly convective stars like HD 61005 are expected to emit winds stronger than the Sun's. The pre-main-sequence M dwarf AU Mic, whose luminosity is even lower than HD 61005's, appears to be emitting such a wind (\citealt{augereau06,schuppler15,chiang17}).

In the AU Mic debris disk and other disks (e.g. HD 35841, \citealt{esposito18}; $\beta$ Pic, \citealt{kalas95}; HD 139664, \citealt{kalas06}), dust needs to be blown onto the wide, barely bound orbits implied by equations (\ref{eq:orbcorr})--(\ref{eq:betamax}) to reproduce their signature halo surface brightness profiles  (\citealt{strubbe06}). Furthermore, \citet{olofsson22} have shown that the spatially extended mm-wave halos discovered by \citet{macgregor18} in HD 61005 and HD 32297 can be explained by the abundance of high-eccentricity grains implied by our top-heavy $\beta$-distribution.

\subsubsection{Scattering phase function (SPF)}
 
We have adopted the Hedman-Stark Saturn G ring SPF which appears consistent with SPFs measured for most debris disks (\citealt{hughes18}), but notably does not match the SPF inferred for the 
debris disk HR 4796A (\citealt{chen20}). Most of the disk morphologies highlighted in this paper, including the ``double wing'' and ``needle'', do not rely on our use of the Hedman-Stark SPF; these shapes would still manifest using other SPFs, e.g.~a Henyey-Greenstein SPF with $g=0.5$ (\citealt{lee16}, their figure 9), modulo order-unity changes in relative brightness. The disk feature most sensitive to SPF is the ``bulb'', which relies on strong forward-scattering; the bulb is seen using Hedman-Stark but not Henyey-Greenstein with $g = 0.5$ (\citealt{lin19}, their figure 3). The HR 4796 SPF has an even stronger forward-scattering peak than the Hedman-Stark SPF, ensuring that bulbs would still be seen using the former.

\section{Comparison with Observations}
\label{sec:compare}

We compare observed scattered light images of individual sources (HD 106906, HD 111520, HD 15115, HD 32297, HD 61005) with model images chosen by-eye to resemble those observations. The model images are drawn from the fiducial models of \S\ref{sec:model}, modified slightly in some cases to attempt a better visual match.  The comparisons are rough as we do not plot observed and model images on a common, absolute surface brightness scale.  However, by ensuring that display scalings (log vs.~sqrt vs.~linear) are the same, and dynamic ranges (max/min pixel ratios) are comparable, we can compare relative brightnesses and thus disk shapes.

\subsection{HD 106906}
\begin{figure*}[ht!]
\plotone{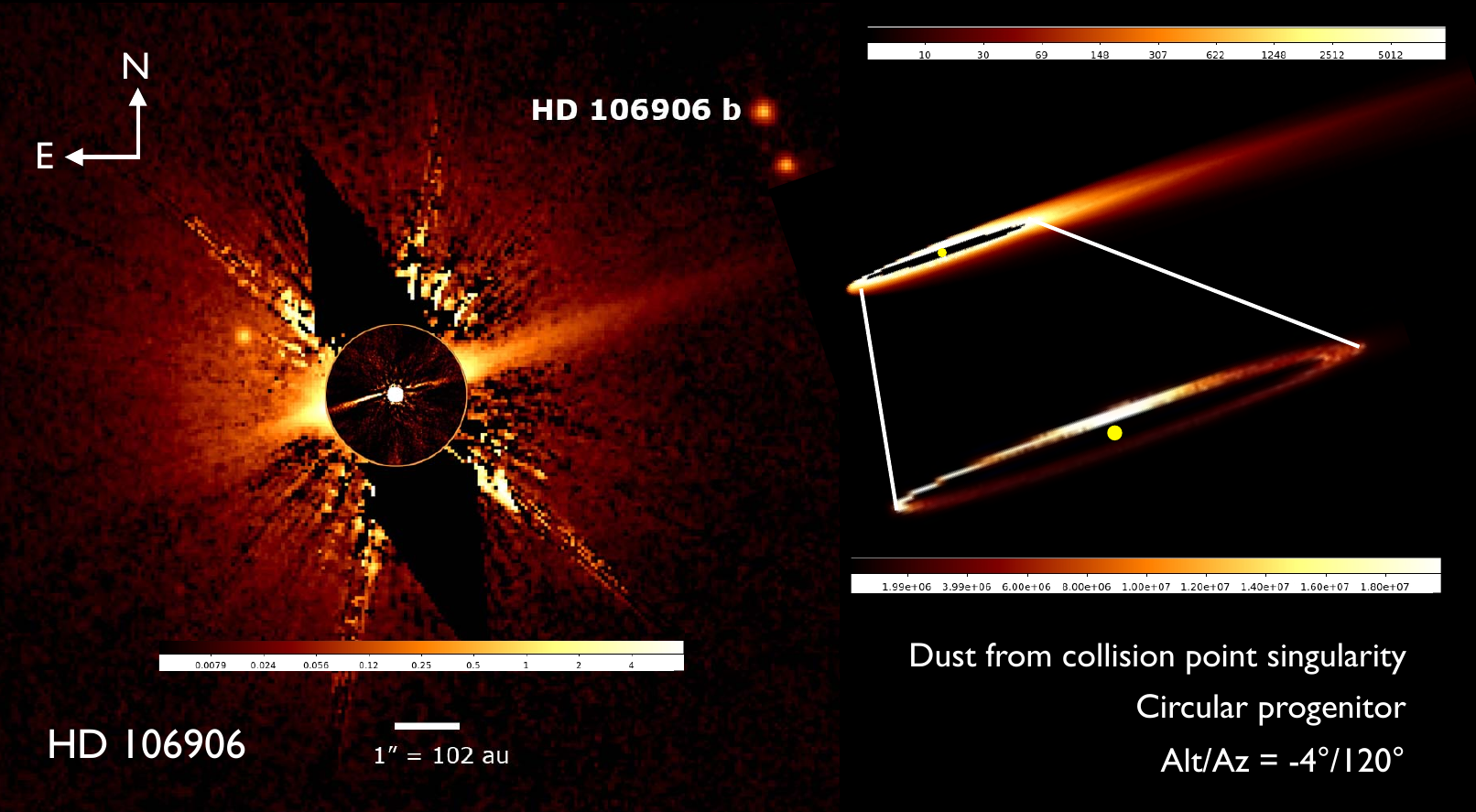}
\caption{{\it Left:} An optical HST/STIS coronagraphic image using the `bb' colormap on a log display stretch ranging up to 8 counts/s/pixel. 
The inset shows an $H$-band  image taken with GPI on a linear stretch between 0 and 1.5 counts/s/pixel \citep{kalas15}. {\it Right:} Model image of dust created from the collision point of a circular progenitor, viewed at Alt/Az = $-4^\circ/120^\circ$ (same as in Fig.~\ref{fig:pano}). The zoom-in on the central cavity should be compared with the GPI inset; it utilizes our mock infrared SPF which is more isotropic than the Hedman-Stark SPF, and a linear brightness scale like the one used in the GPI inset. The number of particles sampled at low $\beta$ has also been increased in the zoom-in to reduce shot noise. The northern half of the model zoomed-in ring, directed toward the observer, is brighter than the southern half because forward-scattering is stronger than back-scattering. Furthermore, the ring's northeast quadrant is brightest because orbital streamlines converge on the collision point situated there and increase line-of-sight column densities. Although these features of our giant impact model match observations, one can also explain the observed asymmetries, and others, by appealing to gravitational perturbations by the known companion HD 106906b \citep{moore23}.
\label{fig:hd106906}}
\end{figure*}

Figure \ref{fig:hd106906}, left panel, shows a composite image of the HD 106906 system in two wavebands. At visible wavelengths (HST/STIS) and large spatial scales, diffuse emission from the nearly edge-on disk extends farther to the west than to the east (\citealt{kalas15}; Esposito et al.~2023, in preparation). At longer infrared wavelengths and smaller spatial scales (Gemini Planet Imager, GPI), 
the western extension is absent, and the central ring instead appears brighter to the east by $\sim$10--30\% \citep{kalas15,crotts21}. 

Together the optical and infrared observations suggest the collision point is to the east, and that the small dust grains created at the collision point are blown onto eccentric orbits whose apastra are to the west. A model scattered light image illustrating this scenario is shown in the right panel of Fig.~\ref{fig:hd106906}; it is the same as the ${\rm Alt}/{\rm Az} = -4^\circ/120^\circ$ image in Fig.~\ref{fig:pano}. Also shown is a zoom-in of the same model image utilizing our mock infrared SPF (\S\ref{subsubsec:proc}), to be compared with the infrared GPI inset. 
The zoomed-in infrared ring is brighter on its forward-scattering northern half, and most bright in its northeast quadrant where orbital streamlines converge on the collision point and line-of-sight column densities are highest (we chose Alt/Az = -4$^\circ$/120$^\circ$ to situate the collision point in this quadrant). For simplicity we have assumed the progenitor to be on a circular orbit, but there is evidence from the GPI data that the host star (actually a binary) is not centered on the inner disk, implying the latter has an eccentricity of $\sim$0.16, possibly higher (\citealt{crotts21}, their figure 7). We could account for this by allowing for an eccentric progenitor; the resultant disk eccentricity could further brighten the northeast quadrant.


What this giant impact scenario does not account for is the $\sim$11 $M_{\rm J}$ companion HD 106906b (Fig.~\ref{fig:hd106906}; see e.g.~\citealt{nguyen21}, and references therein). In principle, HD 106906b can perturb dust grains into a needle-like shape (e.g.~\citealt{lee16}), without the need for a giant impact. \citet[][see also \citealt{rodet17}]{moore23} simulate this dynamics and find that HD 106906b, orbiting exterior to the disk, can excite a coherent disk eccentricity on $> 125$ au scales (their figure 1, bottom panel), and vertically thicken the disk's northwest ansa as observed --- this vertical thickening is absent from our giant impact model which does not account for HD 106906b. \citet{moore23} calculate that continuous secular forcing over the 15 Myr system age would distort the disk out of proportion to the observations (\citealt{crotts21}), and therefore suggest that HD 106906b was only recently captured within the past $\sim$1 Myr. The excess brightness of the inner disk's northeast quadrant would be attributed entirely to the disk's eccentricity, which HD 106906b can marginally excite (\citealt{moore23}, top panel of their figure 1).

\citet{moore23} do not offer model images showing surface brightness asymmetries on scales $> 100$ au, and so the case for HD 106906b being responsible for the scattered light morphology is not complete. For now Occam's Razor would seem to prefer a gravitational perturber scenario over a giant impact.

\subsection{HD 111520}
\begin{figure*}[ht!]
\plotone{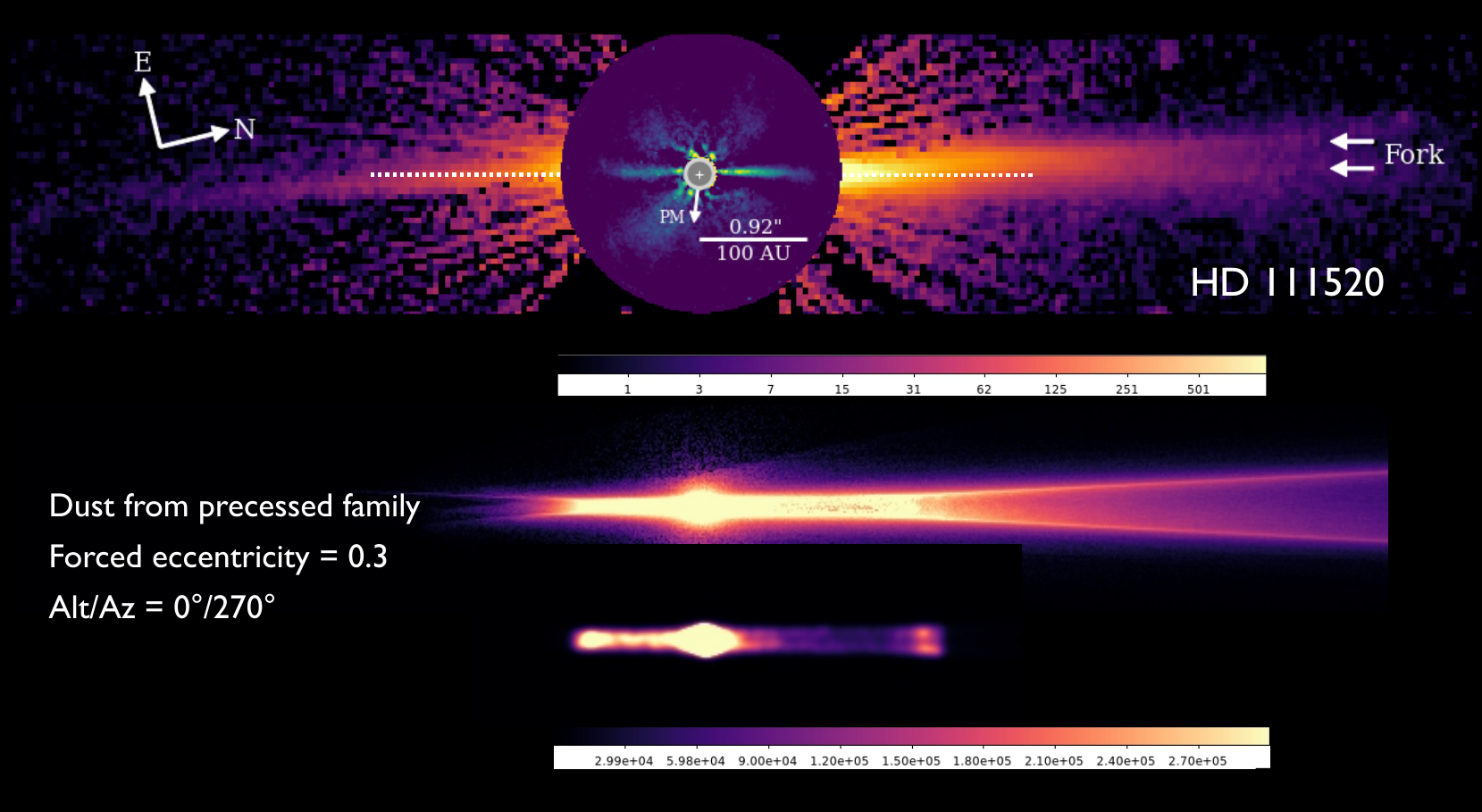}
\caption{{\it Top:} HD 111520 imaged on large scales by HST/STIS (0.59 $\mu$m) and on smaller scales with GPI (1.65 $\mu$m inset), reprinted from figure 4 of \citet{crotts22}. The HST image has a dynamic range in surface brightness of 500 and is displayed on a log scale, while the GPI image is on a linear scale between -1.5 and 10 mJy/arcsec$^2$. The arrow points in the direction of the star's proper motion. {\it Middle:} Model image of dust created from a precessed collisional family with a forced eccentricity of 0.3 and a free inclination of 0.05 rad = 2.9$^\circ$ (same as in Fig.~\ref{fig:precessed_eccentric}). The model image is on a log brightness scale that uses the `magma' color scheme of the \texttt{ds9} imaging tool. About two orders of magnitude in surface brightness are spanned from where the fork first splits to the northern tip; the brightest regions closest to the star are saturated. {\it Bottom:} Same model image on a linear brightness scale that resolves the regions closest to the star, smoothed and better sampled ($N_{\rm p} = 2000$ instead of the usual 200) to reduce shot noise.  On this smaller scale we see a compact eccentric disk that is vertically thinnest at periapse and thickest at apoapse. Whereas this model disk is brighter to the south, the disk observed by GPI is brighter to the north. Also not reproduced by the model is the observation that the GPI disk aligns with only the lower (western) fork, as indicated by the horizontal dotted line in the top panel. In \S\ref{sec:sum} we discuss how, instead of a giant impact, a planet may better explain some of the observed disk features, in a scenario similar to that of $\beta$ Pic b and the warp it induces. 
\label{fig:hd111520}}
\end{figure*}

Figure \ref{fig:hd111520} reprints the composite HST/STIS (0.59 $\mu$m) and GPI (1.65 $\mu$m) image of HD 111520 from \citet[][their figure 4]{crotts22}, alongside an image of dust from one of our fiducial models of a precessed collisional family, having a forced eccentricity of 0.3 and a free inclination of 0.05 rad = 2.9$^\circ$. The model reproduces, superficially, the fork observed in the larger-scale HST image. 
The model fork represents the limb-brightened top and bottom surfaces of a disk whose particles all have the same universal orbital inclination (a few degrees) relative to a reference plane, but whose longitudes of ascending node are randomized. The universal inclination is the free inclination inherited from the progenitor of the collisional family. The randomization of nodes is presumably due to some perturbing mass, perhaps a planet,\footnote{Or multiple planets, or a disk. What matters is the forced inclination vector at the location of the collision family which defines the plane about which family members precess. All masses in the system contribute to this single forced inclination vector \citep{murray_dermott}. Analogous statements apply to the forced eccentricity vector, assumed in our model to have a magnitude of 0.3.} occupying the plane between the upper and lower branches of the fork, on an eccentric orbit whose periapse is to the south. The perturber needs to be eccentric in order to force collisional family members to be similarly eccentric and to be approximately apsidally aligned. In the HST image, the fork is brighter than the regions between the fork by factors ranging up to $\sim$2. This is comparable to the model's fork-interfork brightness contrast, which is a factor of $\sim$2--3; this contrast can be lowered by reducing the free inclination, thereby squeezing the forks together. 

Unfortunately, other details in the observations do not match those in the model. On the small scales probed by GPI (top inset in Fig.~\ref{fig:hd111520}), the disk appears brighter to the north than to the south. Our model on these scales is instead brighter to the south, at the periapse of the eccentric disk; here the disk is vertically thinner and line-of-sight column densities are larger (bottommost image in Fig.~\ref{fig:hd111520}). A second problem is that the disk seen by GPI aligns with only the lower fork---see the horizontal dotted line drawn in Fig.~\ref{fig:hd111520}, and also figure 5 of \citet{crotts22}. From the model, we would instead expect the spine of the GPI disk to cut through the middle of the fork.
A third feature of the HST data not reproduced by the model is that the southern ansa warps downward to the west, by practically the same amount that the upper fork warps upward to the east \citep[][their figure 5]{crotts22}.

These discrepancies would appear to rule out a precessed collisional family as an explanation for the fork. It might still be possible for the lower fork to be interpreted as the giant impact debris streaming to the north from a collision point to the south. But the upper northern fork and its point-symmetry with the southern ansa likely have a different origin. In \S\ref{sec:sum} we suggest that an inclined planet, analogous to $\beta$ Pic b and the warp it induces (e.g.~\citealt{dupuy19}, and references therein), may do a better job of reproducing these features.


\subsection{HD 15115 (The Blue Needle)}\label{subsec:hd15115}

\begin{figure*}[ht!]
\plotone{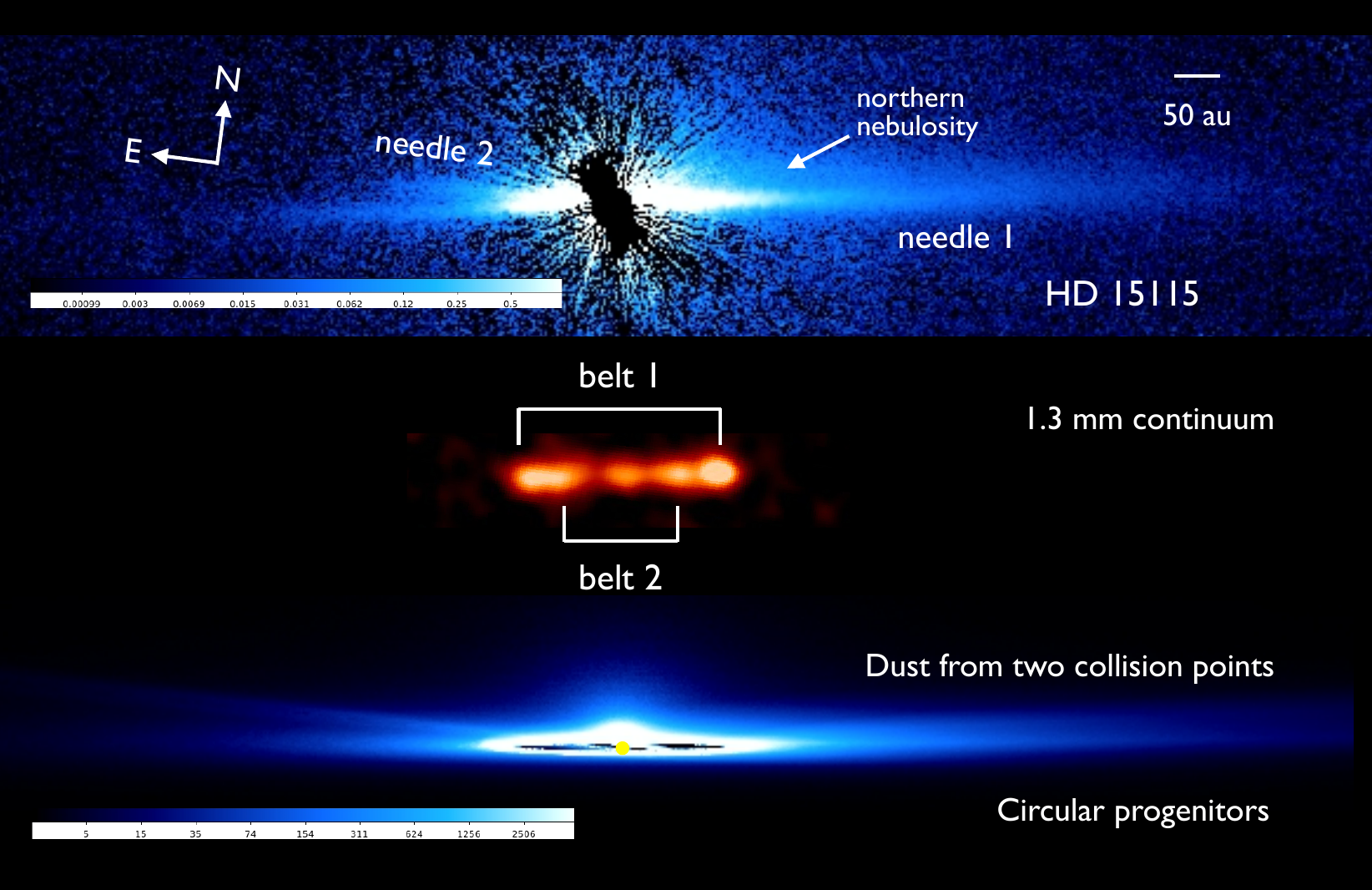}
\caption{{\it Top panel:} HD 15115 imaged by HST/STIS \citep{schneider14} using the `cool' colormap on a log display stretch that ranges up to 1 count/s/pixel. {\it Middle:} Figure 1 of \citet{macgregor19} showing 1.3 mm continuum emission from what appear to be two axisymmetric edge-on parent belts in HD 15115. {\it Bottom:} A model scattered light image with three components: (i) Dust from the collision point of a circular progenitor, viewed at Alt/Az = $-3^\circ/60^\circ$ (cf. Fig.~\ref{fig:pano}). Component (i) features a northern nebulosity resembling that of the HST image. (ii) Dust from the collision point of a second circular progenitor inclined by 6$^\circ$ relative to the first. The second collision point is located at half the orbital radius of the first, and is assumed to produce 1/20 as much dust (by setting $N_{\rm p} = 100$ and $N_{\beta} = 500$ as opposed to our fiducial $N_{\rm p} = 200$ and $N_\beta = 5000$). (iii) Dust from a background circular ring, modeled after \citet{lee16},
containing the first collision point and generating as much as dust as component (i).
\label{fig:hd15115}}
\end{figure*}

HD 15115 is doubly lopsided, a disk that is not only short
to the east and long to the west, but also forked to the east on its short side (Figure
\ref{fig:hd15115}; \citealt{schneider14}). Unfortunately, a short-side-only fork cannot be explained by our precessed collisional family models,
which predict that short-side forks are always accompanied by
 long-side forks which are easier to resolve
(Fig.~\ref{fig:precessed_eccentric}). A long-side fork is not
observed in HD 15115.

An alternative way to characterize the system is to say that it has not one but two needles, a long one pointing west (``needle 1'') and a shorter one pointing almost due east (``needle 2''). 
Unprecessed debris from a collision point can generate a needle (Figs.~\ref{fig:pano} and \ref{fig:pano_ecc}). 
Thus we attempt to reproduce HD 15115 by appealing to the unprecessed debris
from two giant impacts --- one whose collision point lies to the east
to create needle 1, and a second involving a
smaller progenitor disrupted to the west to create needle 2 (Fig.~\ref{fig:hd15115}, bottom panel). The two progenitors would have been on orbits 
mutually inclined by several degrees, with the smaller progenitor (needle 2) colliding
closer to the star so that its near-periastron debris does not protrude from the western arm of needle 1. By suitable choice of Alt/Az the model can also reproduce the ``nebulosity'' to the north of the western arm of needle 1 (see also Fig.~\ref{fig:pano}). 

We have also added to
this mix a
background parent body belt and its axisymmetric halo to better match
the diffuse emission seen farther east and aligned with needle 1. Interestingly, there are two parent body belts in the system, as revealed by ALMA at mm
wavelengths \citep{macgregor19}. Perhaps each parent belt hosts its own giant
impact. Neither of these thermal mm-wave belts evinces a
substantial eccentricity, suggesting that these are background disks
not directly related to the needles seen in scattered light. Dust from the two hypothesized impacts may differ in their mineralogies, in which case we would expect the two needles to differ in their spectroscopic properties.



\subsection{HD 32297}
\begin{figure*}[ht!]
\plotone{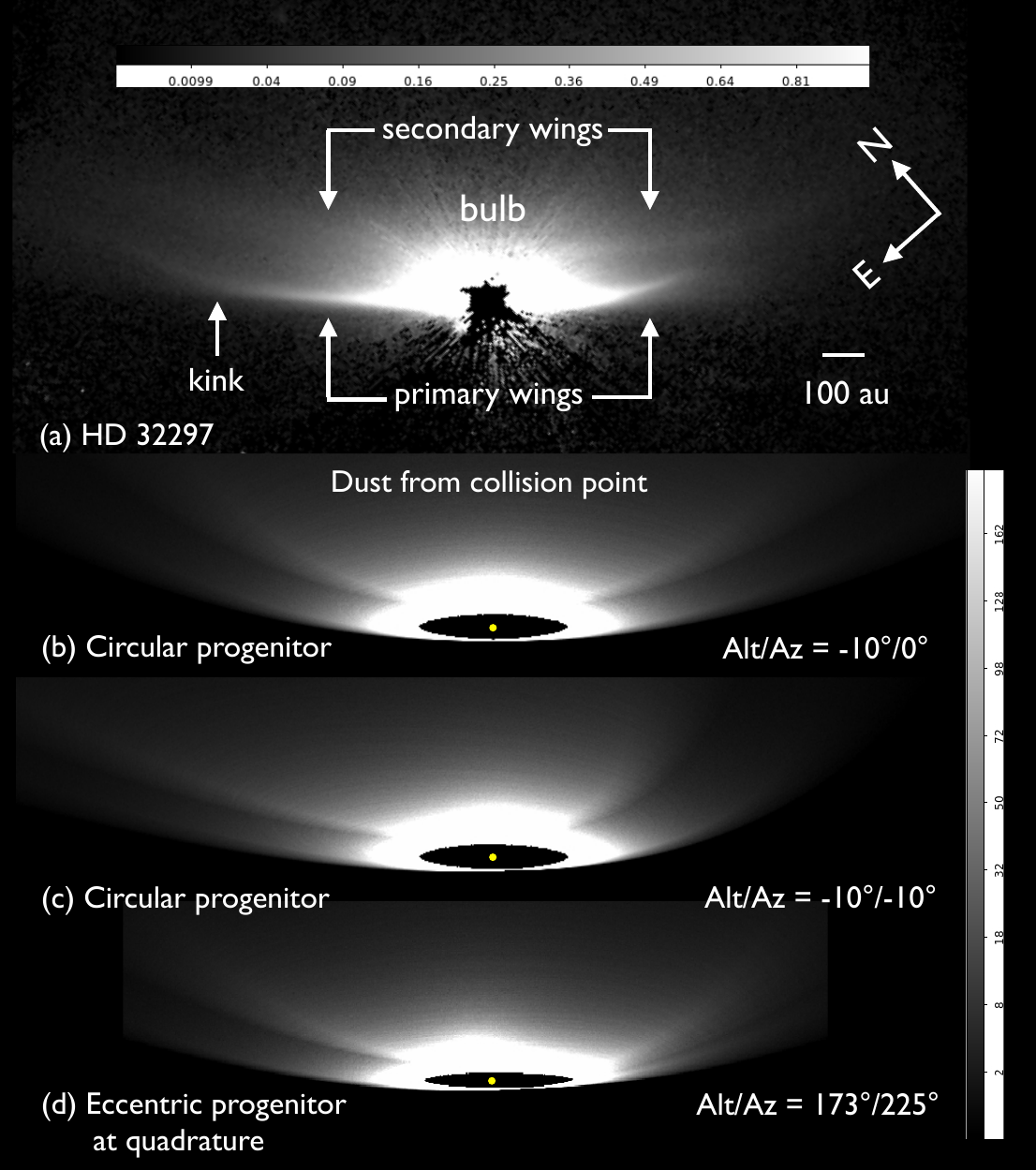}
\caption{(a) HST/STIS image of HD 32297 \citep{schneider14}, on a greyscale square root display from 0 to 1 counts s$^{-1}$ pixel$^{-1}$. (b) Model scattered light image of dust generated from the collision point of a progenitor on a circular orbit, viewed 10$^\circ$ below the dust mean orbital plane and with the collision point located behind the star. The image is derived from the same sequence shown in Fig.~\ref{fig:pano}. (c) Same as the above model image, but turned in azimuth by 10$^\circ$ in an attempt to shorten and tilt upward the right half of the primary wing to better match the HST image. Unfortunately, doing so shortens and upturns even more the right half of the secondary wing, contrary to observation. (d) A model similar to those above, for a progenitor on an eccentricity = 0.4 orbit, derived from the sequence shown in Fig.~\ref{fig:pano_ecc}. The observer lies 7$^\circ$ below the mean dust plane, and the viewing azimuth is chosen to squeeze the primary and secondary wings closer together on the right than on the left, as motivated by the HST image. The match between model and observation is still inadequate, however; the tilts of the wings are off, the secondary wing is too bright compared to the primary, and the left primary wing is not kinked as observed. Also missing from all three model images is an accounting of the scattered light from the massive background disk inferred from ALMA \citep{macgregor18}. All model images are shown on the same greyscale square root display.
\label{fig:hd32297}}
\end{figure*}

Figure \ref{fig:hd32297} compares the HST/STIS image of HD 32297 \citep{schneider14} with model scattered light images of dust created from a singular collision point located behind the star. Viewed nearly edge-on from below the dust plane, the parabola composed of the smallest, most marginally bound grains (\S\ref{subsec:fiducial}) traces out a ``primary wing'' seen over a range of scattering angles, mostly $\theta > 90^\circ$. The jets (\S\ref{subsec:fiducial}) create a ``secondary wing'' seen predominantly in forward-scattered light, $\theta < 90^\circ$. Apastra point nearly directly toward the observer; apastron grains forward-scatter starlight to generate a ``bulb'' (e.g.~\citealt{lin19}). Each of these model features can be identified in the observed HST image, supporting the idea that a giant impact may be responsible.

There are a number of unresolved quantitative issues. 
In the HST image, the primary wing is brighter than the secondary by factors of $\sim$2--3. In our model image using a progenitor on a circular orbit, the brightness contrast between primary and secondary wings is a factor of 1.5--2, lower than desired because the primary tends to be seen in weaker back-scattered light. Ways to remedy this discrepancy include: changing the scattering phase function; allowing for some differential precession of fragment orbits to smear out the collision point, thereby diffusing the jets and by extension the secondary wing; and adding light to the primary wing from a separate background disk. 
Such a background disk must be present based on the strong mm-wave radiation observed by ALMA, emitted by $\sim$$0.57 M_\oplus$ of large particles in a ring with radius $\sim$120 au \citep{macgregor18}. It may be difficult, however, to have the light from this background, presumably circular disk align with that of the primary wing, as the two structures have different radii of curvature. Another challenge is to avoid having the light from the background disk overwhelm the light from the giant impact debris, which presumably derives from a progenitor much less massive than the background disk (\S\ref{sec:hidden}).

Still another problem is the left-right asymmetry in the observations: the southwest wing of the primary is shorter than its northeast counterpart. At the same time, the observed secondary wing does not show such a difference. Our model can generate left-right asymmetries by turning the collision point away from our line of sight to the star (changing Az; Fig.~\ref{fig:pano}), or by having the collision point be located between  periapse and apoapse of an eccentric progenitor orbit (Fig.~\ref{fig:pano_ecc}). But neither scenario seems able to reproduce the particular left-right asymmetry that is observed (Fig.~\ref{fig:hd32297}); for example, changing Az for a circular progenitor can shorten one half of the primary, but not without also shortening half of the secondary, contrary to observation. In the HST image, the northeast half of the primary also ``kinks'' (bends at a sharp angle) at a projected stellocentric distance of $\sim$700 au, in a way that none of our models does.

Do these difficulties rule out a giant impact origin for the double wing? We cannot say with certainty, but feel these discrepancies are less significant than reproducing the double wing in the first place. We suggest that additional physics, notably ram pressure by the interstellar medium, might address the remaining problems (\S\ref{sec:sum}).

\subsection{HD 61005 (The Moth)}
\begin{figure*}[ht!]
\plotone{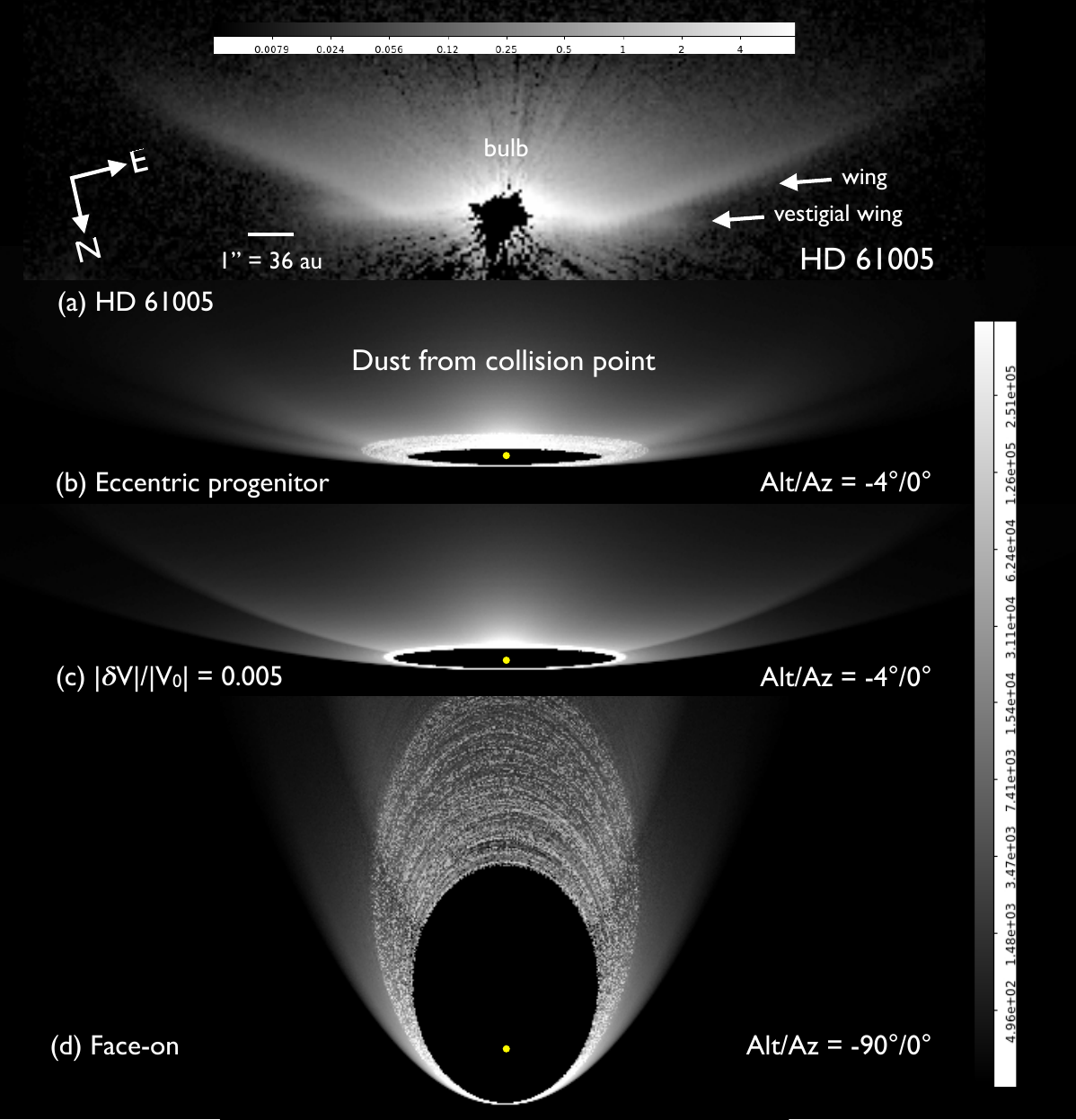}
\caption{(a) HST/STIS image of HD 61005 \citep{schneider14}, on a log greyscale display from 0 to 8 counts s$^{-1}$ pixel$^{-1}$. We can discern, barely, the central cavity of a ring, and the Moth's wings emerging from the ring at an angle of $\sim$23$^\circ$ from the horizontal \citep{buenzli10, esposito16} We also see a central bulb, and a pair of vestigial wings extending $\sim$1'' horizontally past the ring ansae. (b) Model scattered light image of dust from a collision point located at periapse of a progenitor orbit with eccentricity = 0.7, on a log brightness scale as shown. (c) Same as (b), but using $|\delta \mathbf{V}|/|\mathbf{V_0}| = 0.005$ 
 instead of our fiducial 0.025 to produce a sharper image whose parabola-derived wings are shorter than the jet-derived wings. (d) Same model as (b), but viewed face-on and on a log brightness scale that ranges up to $10^5$ in code units instead of $5 \times 10^5$, showing the parabola and jets that project at low $|{\rm Alt}|$ into two pairs of wings (cf.~Figs.~\ref{fig:aftermath_pdf} and \ref{fig:pano}).
\label{fig:hd61005}}
\end{figure*}

To the extent that HD 32297 resembles HD 61005 \citep{schneider14}, we are led to consider a collision point model for the latter as we did for the former, placing the singularity  behind the star (Az $\sim 0$) so that most of the debris is in front and seen nearly but not exactly edge-on (Alt $\sim$ a few degrees). Figure \ref{fig:hd61005} shows how such a model fares against the observations. The model produces the usual primary and secondary wings. The more eccentric the progenitor orbit, and/or the higher the viewing $|{\rm Alt}|$, the more upturned the wings are. We have chosen a progenitor orbital eccentricity of 0.7 and a viewing Alt of -4$^\circ$ to tilt the secondary wings upward by approximately the observed $\sim$23$^\circ$ \citep{buenzli10, esposito16}. This same model can also roughly reproduce the observed central bulb and ring cavity.

The discrepancies between observation and theory in HD 61005 appear more serious than in HD 32297. The primary wings, which derive from the parabola of marginally bound grains (\S\ref{subsec:fiducial}), are much too long compared to the observed ``vestigial'' wings. Another way to characterize this problem is to say that the model brightness ratio between primary and secondary wings differs by more than a factor of 10 from the observed brightness ratio, and with the wrong sense since the observed primary is fainter than the secondary. Furthermore, the secondary model wings, which derive from the jets, are not as straight as the observed Moth wings. Reducing the velocity dispersion $|\delta V_0|/|V_0|$ of fragment ejecta (\S\ref{subsubsec:proc}) dims and shortens the primary wing relative to the secondary, but the effect is modest and inadequate to reproduce the observations. For HD 61005, sculpting by the interstellar medium may be a more dominant effect \citep{debes09,maness09}.


\section{A Hidden Population of Giant Impact Progenitors}\label{sec:hidden}


We have not computed the progenitor masses implicated in our candidate giant-impact hosting disks. We reserve that work for a future study that replaces our code units with physical units to place our synthetic images on an absolute surface brightness scale, for comparison with similarly calibrated observations. Here we outline some general considerations regarding progenitor mass, and how many hidden progenitors are implied by the observed destruction of a few.

Catastrophic disruption of a progenitor disperses its  fragments onto separate orbits, with no fragment larger than half the size of the progenitor. On the scales of interest here, catastrophic dispersal seems possible for progenitors up to Pluto in size: if the colliders are drawn from a reservoir at 100 au from a $2 M_\odot$ star with an eccentricity/inclination dispersion of order 0.3 (similar to the dynamically hot and more massive component of the Kuiper belt), collision velocities (gravitationally unfocussed) are $\sim$1 km/s, comparable to the surface escape velocity from Pluto.

If bodies larger than Pluto collide, only a fraction of their mass would be dispersed as debris. The collision would be gravitationally focussed, occurring nearly head-on at a velocity comparable to the bodies' mutual surface escape velocity. Such a collision has high accretion efficiency, with at most $\sim$10\% of the mass escaping as debris from the gravitational well of the merger product (see, e.g., figure 5 of \citealt{emsenhuber20}).

Statistically, observing a collision between two bodies implies the existence of a reservoir of similarly sized bodies. Consider $N$ bodies with collision cross section $\sigma$ and relative velocity $v_{\rm rel}$ occupying a volume $V$ around a star. Say the debris of a pairwise collision is observable for time $t_{\rm obs}$. Then the probability of observing the debris from one collision is
\begin{equation}
P_{\rm obs} = n^2 \sigma v_{\rm rel} V t_{\rm obs}  \label{eqn:single}
\end{equation}
where $n = N/V$ is the number density. The volume $V$ may be approximated as that of a torus of circumference $2\pi a$, radial width $v_{\rm rel}/\Omega$, and vertical height $v_{\rm rel}/\Omega$, for orbital radius $a$ and angular frequency $\Omega$, assuming relative velocities $v_{\rm rel}$ are close to isotropic. We further assume gravitationally unfocused collisions between spheres of radius $R$ so that $\sigma = \pi (2R)^2$. Then dropping order-unity constants, we may solve for $N$:
\begin{equation}
N \sim \frac{a}{R} \sqrt{\frac{1}{\Omega t_{\rm obs}}}
\sqrt{\frac{v_{\rm rel}}{\Omega a}} \sqrt{P_{\rm obs}} \,.
\end{equation}
In the context of our paper, the duration of observability $t_{\rm obs}$ is the time interval over which the giant impact debris generates a detectable non-axisymmetric morphology. We do not have a precise measure of $t_{\rm obs}$. Kral et al.~(\citeyear{kral15}; see also \citealt{lohne08}) provide estimates for how long the debris from a giant impact remains asymmetric, assuming the debris collides only with itself. The asymmetric phase may instead be limited by how the giant impact debris is ground away by bodies in a background parent belt \citep{chiang17}. We do not dwell on details here but argue roughly as follows. An upper limit on $t_{\rm obs}$ is the system age $t_{\rm age} \sim 10$--100 Myr. Unless we are observing the system at some special moment, $t_{\rm age}$ should also be the collisional lifetime (the time to grind away half the mass) of the background parent belt of dust mass $M_{\rm bkgd}$, assumed to be $\gtrsim M_{\rm coll}$, the mass of the giant impact debris.
 Then since collisional lifetimes scale as $1/M$, we have
\begin{equation}
t_{\rm obs} \sim \frac{M_{\rm coll}}{M_{\rm bkgd}} \,t_{\rm age} \,.
\end{equation}
Suppose $M_{\rm bkgd} \sim 0.01$--$1 M_\oplus$ (the range of mm-wave masses inferred for disks including, e.g., AU Mic, HD 15115, and HD 32297; \citealt{matthews15,panic13,macgregor18}), and $M_{\rm coll} \sim 0.001 M_\oplus$ (comparable to Pluto). Then $M_{\rm coll}/M_{\rm bkgd} \sim 10^{-1}$--$10^{-3}$, $t_{\rm obs} \sim 10^4$--$10^7$ yr, and
\begin{align}
N & \sim  2 \times 10^4 \left( \frac{a}{100 \, {\rm au}} \right)^{7/4} \left( \frac{1000 \, {\rm km}}{R} \right) \left( \frac{10^6 \, {\rm yr}}{t_{\rm obs}} \right)^{1/2} \nonumber \\
&  \times \left( \frac{v_{\rm rel}/\Omega a}{0.1} \right)^{1/2} \left( \frac{P_{\rm obs}}{0.1} \right)^{1/2} \label{eqn:number}
\end{align}
for a total reservoir mass of
\begin{align} \label{eqn:mass}
M_{\rm reservoir} & =  N M_{\rm coll} \sim 20 M_{\oplus} \left( \frac{a}{100 \, {\rm au}}  \right)^{7/4} \nonumber \\
& \times \left( \frac{R}{1000 \, {\rm km}} \right)^2 \left( \frac{10^6 \, {\rm yr}}{t_{\rm obs}} \right)^{1/2}  \nonumber \\
& \times \left( \frac{v_{\rm rel}/\Omega a}{0.1} \right)^{1/2} \left( \frac{P_{\rm obs}}{0.1} \right)^{1/2}\,.
\end{align}
The detection probability $P_{\rm obs}$ equals the fraction of debris disks (assumed to be of comparable age) that show evidence for a recent giant impact.  
This fraction is highly uncertain and our nominal estimate of 0.1 is just a placeholder. Still, our answers for $N$ and $M_{\rm reservoir}$ are not very sensitive to $P_{\rm obs}$.  

An investment of $\sim$$20 M_\oplus$ in $R \sim 1000$ km, Pluto-sized objects would be astounding. For comparison, in the coagulation simulations of Kenyon \& Bromley (e.g.~\citeyear{kenyon12}, and references therein), the number of 1000-km objects formed after 300 Myr at $a \sim 30$ au under minimum-mass solar system disk conditions is $\sim$20 (see their figure 2), about two orders of magnitude lower than equation (\ref{eqn:number}) suggests (after scaling for the smaller $a$). The numbers quickly become more prosaic, however, if we consider the destruction of smaller bodies. If the disruption of $R \sim 100$ km objects is observable, then using this value for $R$ in equations (\ref{eqn:number}) and (\ref{eqn:mass}) with all other variables held fixed at their nominal values would predict $N \sim 2 \times 10^5$ and $M_{\rm reservoir} \sim 0.2 M_\oplus$, values more in line with \citet{kenyon12}.

\begin{deluxetable*}{cccccc}


 \tablecaption{Possible Explanations for Debris Disk Scattered-Light Asymmetries \label{tab:sample}}

 \tablehead{
 \colhead{Disk ID} & \colhead{Giant Impact} & \colhead{ISM Sculpting} & \colhead{Gravitational Perturber} & \colhead{Comments} & \colhead{References} }

 \startdata 
 HD 15115 & Y &  &  & double needle = double impact & 1, 2 \\
  AU Mic & Y & & & collision point avalanches & 3 \\
 HD 32297 & Y & ? & & double wings & 1, 4, 5 \\
 HD 61005 & ? & Y & & straight wings + vestigial wings & 1, 4, 5, 6 \\
 $\beta$ Pic & Y & & Y & needle from giant impact; warp from $\beta$ Pic b & 7, 8, 9, 10 \\
  HD 106906 & Y &  & Y & Occam's razor prefers HD 106906b & 1, 11, 12, 13 \\
 HD 111520 &  &  & Y & warp analogous to $\beta$ Pic warp & 1, 9 \\
 \enddata

\tablecomments{Y = Yes, this seems a viable explanation. ? = Possibly relevant but probably not a dominant effect.}
\tablerefs{1.~This paper 
2.~\cite{mazoyer14} 
3.~\cite{chiang17} 4.~\cite{debes09}
5.~\cite{maness09} 
6.~\cite{olofsson16} 
7.~\cite{dent14} 8.~\cite{janson21} 9.~\cite{mouillet97} 10.~\cite{dupuy19} 
11.~\cite{lee16} 
12.~\cite{rodet17}
13.~\cite{moore23}
}

\end{deluxetable*}

\section{Summary}
\label{sec:sum}

Giant impacts are candidates for explaining non-axisymmetries in debris disks. 
We have computed scattered light images of the debris left by giant impacts in two cases. In case one, dust emerges geyser-like from the ``collision point'' marking the location of the original impact \citep{jackson14}. In case two, the orbits of collision fragments are differentially precessed by a perturbing mass (as asteroid families are perturbed by solar system planets), and collisions between those precessed fragments create dust over a range of orbital azimuths. In both cases, the quasi-steady dust size distribution diverges toward the radiation+wind blow-out limit, preferentially populating highly elliptical, near parabolic ``halo'' orbits \citep{strubbe06}.
The halo orbits are most apsidally coherent (with a well-defined apse at every semi-major axis) when originating from a single collision point (case one), giving rise in scattered light to ``double wings'', ``needles'', and ``bulbs'' depending on the viewing orientation. When the halo orbits derive instead from collision fragments precessed by a perturber (case two), they can generate ``forks'', akin to the zodiacal dust bands from asteroid collision families \citep{dermott84}. Left-right asymmetries in surface brightness abound in both cases, including for collision progenitors on circular orbits.

Though non-axisymmetry is the norm in the aftermath of a giant impact, some of the above morphologies are not unique to giant impacts. Bulbs are generically produced by scattering phase functions that strongly forward scatter \citep{lin19}, and needles can also result from disks shaped by eccentric planets \citep{lee16}. By contrast, the double wing requires a degree of apsidal alignment that seems difficult to achieve except through a giant impact. The double wing is seen when the collision point is situated nearly but not exactly behind the star relative to the observer. One set of wings consists of the arms of the parabolic orbit traced by the most marginally bound grains. A second pair of wings is composed of grains on smaller orbits, lingering at their apastra which point toward the observer. Both sets of wings are seen in forward-scattered starlight.

We also generated thermal emission maps of dust from a collision point, suitable for comparison with ALMA or JWST observations. In nearly edge-on views of the disk, the collision point shows up as a factor-of-2 local over-brightness. The brightening  near the collision point arises because of a local,  order-unity enhancement in the line-of-sight column density.

We compared our model images with  observations of five debris disks with strongly non-axisymmetric shapes. Table \ref{tab:sample} summarizes our views on whether the scattered-light morphologies of these systems are best explained by giant impacts, sculpting from the interstellar medium (ISM), or perturbations from a nearby massive body.

HD 15115 presents perhaps the clearest case for impacts. The system features two needles, one pointing west and a shorter one pointing east. The hypothesized giant impact underlying each needle throws dust onto orbits made highly elliptical by stellar radiation+wind pressure, their apastra pointing down the needle's long end. As the two needles differ in brightness and the locations of their collision points, they trace different collision circumstances with potentially different dust mineralogies. Detecting spectroscopic differences between the two needles could support this hypothesis; see, e.g., \citet{telesco05} for observations along these lines for $\beta$ Pic. 

HD 32297 and HD 61005 both feature, to different extents, double wings. Our collision-point model reproduces the gross appearance of HD 32297 in scattered light. However, we cannot account for the optical disk's particular left-right asymmetries, which might owe their origin instead to drag from the interstellar medium (ISM; \citealt{scherer00}; \citealt{debes09}; \citealt{maness09}). HD 61005 may be more if not entirely dominated by ISM sculpting, as a giant-impact-only model fails to reproduce how much brighter, and how straight, its dominant pair of wings are. Current ISM models have calculated the effects of unidirectional ram pressure on individual grain orbits, but have yet to treat how an ensemble of grains responds, i.e.~how secular forcing by gas, ram-pressure blow out, and grain-grain collisions work together to shape the grain size and orbital distributions. How the ISM interacts with any host stellar wind may also be significant. 

HD 106906 presents a needle morphology, together with a brightening on the short end of its needle that could be interpreted as the collision point. But these features are also potentially explained by gravitational perturbations from the known companion HD 106906b \citep{lee16,rodet17,moore23}. The companion has the added advantage of being able to vertically thicken the disk as observed, depending on its formation history \citep{moore23}. Scattered light images on scales $> 100$ au need to be generated with the parameters of HD 106906b to further support the case of gravitational perturbations over a giant impact.

HD 111520 is needle-shaped and forked, but in a way that our giant impact models cannot explain. In particular, a portion of the disk is warped, with dust at small radii lying in one orbital plane, and at large radii lying in another \citep{crotts22}. This warp is reminiscent of the one in the $\beta$ Pic debris disk, for which the leading explanation is secular gravitational forcing by the giant planet $\beta$ Pic b (e.g., \citealt{mouillet97,heap00,lagrange10,lagrange12,millarblanchaer15,dupuy19}). Applying the same explanation to HD 111520 would predict a planet lying in the inner disk plane. Further positing that this planet occupies an eccentric orbit might help to explain the system's overall lopsidedness \citep{lee16}.

Other disks that we have not modeled explicitly and where giant impacts are thought to have occurred on Kuiper-belt-like scales include AU Mic and $\beta$ Pic. In AU Mic, a collision point at an orbital radius of $\sim$35 au, within the system's main belt, is thought to be where dust avalanches are triggered by a gusty stellar wind \citep{chiang17}.\footnote{The \citet{chiang17} model to explain time variability in AU Mic is sometimes labeled a `two-ring' model, but it is better described as a giant impact model. The hypothesized secondary ring, distinct from the more massive primary `background' ring, is composed of the fragments of a Varuna-sized collision progenitor.} Each avalanche produces a discrete cloud of dust which disk rotation sends to the southeast, as observed \citep{boccaletti15}; the clouds sighted farthest northwest \citep{boccaletti18} suggest the collision point lies $\sim$0.5'' northwest of the star, and indeed the disk is brighter there than to the southeast (see figure 38 of \citealt{schneider14}, middle panel). The progenitor that created the collision point in AU Mic is estimated to be several hundred km in diameter, similar in size to the large Kuiper belt object Varuna.  

In $\beta$ Pic, a collision point is posited $\sim$80 au from the star to the southwest, where overdensities are observed in the sub-mm continuum and CO \citep{dent14} and in atomic carbon \citep{cataldi18}. A collision point to the southwest predicts a needle pointing northeast, and indeed such a feature is present in $\beta$ Pic on $\sim$1000 au scales (\citealt{larwood01}, their figure 1; \citealt{janson21}). The progenitor mass needed to reproduce the sub-mm overdensity is estimated to be Mars-sized \citep{dent14}. \citet{cataldi18} offer an alternative scenario which replaces a giant impact with a tidal disruption event, and  estimate the disrupted mass to be $\gtrsim 3$ lunar masses.  Still another view comes from \citet{telesco05}, who calculate that the destruction of a planetesimal more than 100 km in diameter is responsible for the mid-infrared clump observed $\sim$52 au to the southwest of the star. These various calculations have yet to be reconciled. 


Overall, the evidence for catastrophic disruption of dwarf planets in the outermost reaches of debris disks, though still tentative, is growing. Further strengthening the case would seem worthwhile as it offers a way to probe otherwise invisible masses (\S\ref{sec:hidden}).

\vspace{0.75in}
Data and codes are available upon request from the authors. Animations of Figures \ref{fig:pano} and \ref{fig:precessed_eccentric} are available at  \url{https://github.com/joshwajones/jones_etal_animations/}. 
We thank Haiyang Wang for discussions on the problem of apsidal alignment, Jonathan Lin for helping us navigate Eve Lee's light scattering code, Katie Crotts and Brenda Matthews for information relating to observations of HD 111520, and the Jena debris disk conference organizers and participants, including Christine Chen, Mohammed Farhat, Yinuo Han, Sasha Krivov, Elisabeth Matthews, Julien Milli, Johan Olofsson, Tim Pearce, Kate Su, and Mark Wyatt, for useful and encouraging exchanges. Meredith Hughes and Rebekah Dawson gave feedback on a draft that led to new thermal emission maps and a revised discussion of HD 106906. An anonymous referee provided an extensive and thoughtful report that motivated many other changes. This work was supported by HST-AR-16608.001-A and HST-GO-15653 from the Space Telescope Science Institute under NASA contract NAS5-26555, and a  Berkeley Discover award supporting undergraduate research in the Physics \& Astronomy departments.

\bibliography{debris}

\appendix
\label{appendix}

We list here the equations used to compute the Keplerian orbital elements of dust grains starting from the Cartesian positions $\{{\mathbf R}\}$ and velocities $\{ {\mathbf V} \}$ of parent bodies. The equations are standard and published elsewhere (see, e.g., the teaching memorandum series of \url{www.rene-schwarz.com}) but we compile them here for convenience.

For a given $\mathbf{R}$ and $\mathbf{V}$ we compute 
the semimajor axis as
\begin{equation}
a = \left( \frac{2}{R} - \frac{V^2}{GM_\odot} \right)^{-1} \nonumber
\end{equation}
where $R = |\mathbf{R}|$, $V = |\mathbf{V}|$, $G$ is the gravitational constant, and $M_\odot$ is the assumed mass of the host star. From the specific angular momentum
\begin{equation}
\mathbf{h} = \mathbf{R} \times \mathbf{V} \nonumber
\end{equation}
we compute the eccentricity vector
\begin{equation}
\mathbf{e} = \frac{\mathbf{V} \times \mathbf{h}}{GM_\odot} - \frac{\mathbf{R}}{R} \,. \nonumber
\end{equation}
The eccentricity $e = |\mathbf{e}|$, and the true anomaly
\begin{equation}
f = \begin{cases}
\arccos [{\mathbf e} \cdot {\mathbf R}/(eR)] & {\rm if} \,{\mathbf R} \cdot {\mathbf V} \geq 0 \\
2\pi - \arccos [{\mathbf e} \cdot {\mathbf R}/(eR)]  & {\rm otherwise}. \nonumber
\end{cases}
\end{equation}
The inclination
\begin{equation} \nonumber
i = \arccos (h_z / h)
\end{equation}
where $h_z$ is the $z$-component of ${\mathbf h}$, and $h = |{\mathbf h}|$. We use a vector pointing from the star to the ascending node 
\begin{equation} \nonumber
\mathbf{n} = \mathbf{\hat{z}} \times \mathbf{h}
\end{equation}
to evaluate the longitude of ascending node
\begin{equation} \nonumber
\Omega = \begin{cases}
\arccos (n_x / n) & {\rm for} \, n_y \geq 0 \\
2\pi - \arccos (n_x / n) & {\rm for} \, n_y < 0
\end{cases}
\end{equation}
with $n_x$ and $n_y$ the $x$ and $y$ components of ${\mathbf n}$, respectively, and $n = |\mathbf{n}|$. Finally, the argument of periastron is given by
\begin{equation} \nonumber
\omega = \begin{cases}
\arccos [{\mathbf e} \cdot {\mathbf n}/(en)] & {\rm if} \,e_z \geq 0 \\
2\pi - \arccos [{\mathbf e} \cdot {\mathbf n}/(en)]  & {\rm if} \, e_z < 0 
\end{cases}
\end{equation}
with $e_z$ the $z$-component of ${\mathbf e}$. 

Parent bodies with unprimed orbital elements are mapped to dust particles with primed elements using
\begin{align}
\nonumber
a' &= \frac{a (1-e^2) (1-\beta)}{1-e^2-2\beta(1+e\cos f)} \\
\nonumber
e' &= \frac{\sqrt{e^2+2\beta e \cos f + \beta^2}}{1-\beta} \\
\nonumber
\omega' &= \omega + \arctan\left(\frac{\beta\sin f}{e + \beta\cos f} \right)  \\
\nonumber
i' &= i \\
\nonumber
\Omega' &= \Omega
\end{align}
where $\beta$ is the force ratio between stellar radiation pressure and stellar gravity. The mean anomaly $M'$ is drawn as a uniform deviate from 0 to $2\pi$ and combined with $e'$ to solve for $f'$ using Kepler's equation. The above formula for semimajor axis $a'$ leads to the scaling with orbital period $\propto (a')^{3/2}$ in equation (\ref{eq:orbcorr}) for the halo grain size distribution, with variations in $a (1-e^2)$ ignored.

\end{document}